\documentclass[%
reprint,
showkeys,
showpacs,
nofootinbib,
amsmath,amssymb,
aps,
floatfix,
10pt]{revtex4-1}
\usepackage{graphicx}% Include figure files
\usepackage{dcolumn}% Align table columns on decimal point
\usepackage{bm}% bold math
\usepackage[colorlinks,citecolor=blue,urlcolor=blue,linkcolor=blue]{hyperref}% add hypertext capabilities
\usepackage[figtopcap]{subfigure}
\begin{document}
%\preprint{CTP/W2-9-14-fT}
%%%%%%%%%%%%%%%%%%%%%%%%%%%%%%%%%%%%%%%%%%%%%%%%%%%%%%%%%%%%%%%%%%%%%%%%%%%%%%%%%%%%%%
\title{The Hidden Flat Like Universe II\\
Quasi inverse power law inflation by $f(T)$ gravity}
\author{W. El Hanafy$^{1,3}$}%
\email{waleed.elhanafy@bue.edu.eg}
\author{G.G.L. Nashed$^{1,2,3}$}
\email{nashed@bue.edu.eg}
\affiliation{$^{1}$Centre for theoretical physics, the British University in Egypt, 11837 - P.O. Box 43, Egypt.}
\affiliation{$^{2}$Mathematics Department, Faculty of Science, Ain Shams University, Cairo, Egypt.}
\affiliation{$^{3}$Egyptian Relativity Group (ERG).}
%\date{\today}% It is always \today, today,
             %  but any date may be explicitly specified
%%%%%%%%%%%%%%%%%%%%%%%%%%%%%%%%%%% Abstract %%%%%%%%%%%%%%%%%%%%%%%%%%%%%%%%%%%%%%%%
\begin{abstract}
In a recent work, a particular class of $f(T)$ gravity, where $T$ is the teleparallel torsion scalar, has been derived. This class has been identified by flat-like universe (FLU) assumptions \cite{HN15}. The model is consistent with the early cosmic inflation epoch. A quintessence potential has been constructed from the FLU $f(T)$-gravity. We show that the first order potential of the induced quintessence is a quasi inverse power law inflation with an additional constant providing an end of the inflation with no need to an extra mechanism. At $e$-folds $N_{*}= 55$ before the end of the inflation, this type of potential can perform both $E$ and $B$ modes of the cosmic microwave background (CMB) polarization pattern.
\end{abstract}

\pacs{98.80.-k, 04.50.Kd, 98.80.Cq, 98.80.Es}% PACS, the Physics and Astronomy
                             % Classification Scheme.
\keywords{modified gravity, early universe, cosmology}%Use showkeys class option if keyword
                              %display desired
\maketitle
%\tableofcontents
%%%%%%%%%%%%%%%%%%%%%%%%%%%%%%%%%%%%%%%%%%%%%%%%%%%%%%%%%%%%%%%%%%%%%%%%%%%%
\section{Introduction}\label{Sec1}
%%%%%%%%%%%%%%%%%%%%%%%%%%%%%%%%%%%%%%%%%%%%%%%%%%%%%%%%%%%%%%%%%%%%%%%%%%%%
Inflation currently represents a leading frame exploring possible overlaps between gravity and quantum field theory and explaining the initial conditions of our universe. Cosmic inflation is a very early accelerating phase usually represented by an exponential expansion just $\sim 10^{-35}$ sec after the Big Bang. As a result, the universe becomes an isotropic, homogeneous and approximately flat. Standard inflation assumes existence of an inflaton (scalar) field, whose potential governs the inflation model. This implies a special treatment of the scalar field on a curved spacetime background. During this stage when the initial quantum fluctuations cross the horizon it transforms into classical fluctuations. In the acceptable inflation models, the spectrum of the produced fluctuations are tiny deviated from being a scale invariant. The deviations from the scale invariant spectrum are according to the considered potential. At the end of the inflation when the inflaton potential drops to its effective minimum, it allows the scalar field to decay through a reheating process restoring the big bang nucleosynthesis epoch. During this stage the primordial fluctuations transform into photon and matter fluctuations. Later this causes the CMB anisotropy, which has an important impact on the structure formation at later stages.

As is well known, the primary CMB temperature signal is snapshot of acoustic oscillations at recombination, i.e. red-shift $z \sim 1100$. The CMB polarization pattern can be decomposed into two components: (i) Curl-free (gradient-mode) component, called $E$-mode (electric-field like), generated by both the scalar and tensor perturbations at recombination and reionization. (ii) Grad-free (curl-mode) component, called $B$-mode (magnetic-field like), generated by vector or tensor perturbations, e.g by gravitational waves from inflation.

Recent measurements of the fluctuations by the Planck satellite \cite{Pl1, Pl2} and the BICEP2 experiment \cite{BICEP2} provide good constrains on the assumed inflation potentials, going backwards in time allowing to choose the right initial inflationary potential. These measurements may define two inflation observable parameters, the spectral scalar index (scalar tilt) $n_{s}$ and the tensor-to-scalar ratio $r$. Some models predict only $E$-modes so that their tensor-to-scalar ratios having small values, while others predict $B$-modes so they having large tensor-to-scalar ratios.

Also, inflation has been treated within modified gravity theories framework such as $F(R)$ gravity. The $F(R)$ action in Jordan frame transforms into Einstein-Hilbert action plus scalar field in Einstein frame by means of a conformal transformation. This allows to define a scalar field potential in terms of the $F(R)$ theory under consideration. In the $F(T)$ theories, where $T$ is teleparallel torsion scalar, the case is different due the lack of the invariance under conformal transformation. In a recent work, we have developed an alternative technique by using a semi-symmetric torsion in $F(T)$ cosmic applications \cite{HN15}. This allows to map the torsion contribution in the modified Friedmann equations into a scalar field.

The organization of the work can be presented as follows: In Section \ref{Sec2}, we define the used notations and the $F(T)$ gravity theories in addition to a summary on a particular class of $F(T)$ gravity motivated by FLU assumptions. In Section \ref{Sec3}, we design an initial inflation potential capable to perform double tensor-to-scalar ratios for a single scalar tilt parameter. We suggest the obtained potential to perform both $E$ and $B$ modes of the CMB polarization. In Section \ref{Sec4}, we consider the case of the semi-symmetric torsion tensor, when the torsion potential is made of a scalar field. In a previous work, we have studied the simplest case of the obtained potential. We extend our investigation, here, to include higher order effects of the potential. The model predicts a quasi inverse power law inflation with a graceful exit with no need to an additional mechanism. A more interesting results are obtained by calculating the slow roll parameters of this model showing its predictions of the inflation observable parameters. Finally, the work is concluded in Section \ref{Sec5}.
%%%%%%%%%%%%%%%%%%%%%%%%%%%%%%%%%%%%%%%%%%%%%%%%%%%%%%%%%%%%%%%%%%%%%%%%%%%%
\section{Notations and Background}\label{Sec2}
%%%%%%%%%%%%%%%%%%%%%%%%%%%%%%%%%%%%%%%%%%%%%%%%%%%%%%%%%%%%%%%%%%%%%%%%%%%%
Teleparallel geometry has provided a new scope to examine gravity. Within this geometry a theory of gravity equivalent to the general relativity (GR), the teleparallel theory of general relativity (TEGR), has been formulated \cite{M2002}. The theory introduced a new invariant (teleparallel torsion scalar) constructed from the torsion tensor instead of Ricci invariant in the Einstein-Hilbert action. Although, the two theories are equivalent on the field equations' level, they have different qualities on their Lagrangian's level. Some applications in astrophysics are given in \cite{nashed:1996,nashed:2010,nashed:2003,nashed:1997,nashed:2002,nashed:2006,nashed:2007}. We show below the main construction of the teleparallel geometry.
\subsection{Teleparallel space}\label{Sec2.1}
In this section we give first a brief account of the absolute parallelism (AP)-space. This space is denoted in the literature by many names teleparallel, distant parallelism, Weitzen\"{o}ck, absolute parallelism, vielbein, parallelizable space.  Recent versions of vielbein space with a Finslerian flavor may have an important impact on physical applications \cite{Wanas2,Nabil1,Nabil2,Nabil3}. An AP-space is a pair $(M,\,h_{a})$, where $M$ is an $n$-dimensional smooth manifold and $h_{a}$ ($a=1,\cdots, n$) are $n$ independent vector fields defined globally on $M$. The vector fields $h_{a}$ are called the parallelization vector fields.

Let\, $h_{a}{^{\mu}}$ $(\mu = 1, ..., n)$ be the coordinate components of the $a$-th vector field $h_{a}$. Both Greek (world) and Latin (mesh) indices are constrained by the Einstein summation convention. The covariant components $h_{a \mu}$ of $h_{a}$ are given via the relations
\begin{equation}\label{orthonormality}
h_{a}{^{\mu}}h^{a}{_{\nu}}=\delta^{\mu}_{\nu}\quad \textmd{and}\quad h_{a}{^{\mu}}h^{b}{_{\mu}}=\delta^{b}_{a},
\end{equation}
where $\delta$ is the Kronecker tensor. Because of the independence of $h_{a}$, the determinant $h:=\det (h_{a}{^{\mu}})$ is nonzero.

However, the vielbein space is equipped with many connections \cite{Wanas1,Wanas3,Wanas4,Waleed4}, on a teleparallel space $(M,\,h_{a})$, there exists a unique linear connection, namely Weitzenb\"{o}ck connection, with respect to which the parallelization vector fields $h_{a}$ are parallel. This connection is given by
\begin{equation}\label{W_connection}
\Gamma^{\alpha}{_{\mu\nu}}:=h_{a}{^{\alpha}}\partial_{\nu}h^{a}{_{\mu}}=-h^{a}{_{\mu}}\partial_{\nu}h_{a}{^{\alpha}},
\end{equation}
and is characterized by the property that
\begin{equation}\label{AP_condition}
\nabla^{(\Gamma)}_{\nu}h_{a}{^{\mu}}:=\partial_{\nu}
{h_a}^\mu+{\Gamma^\mu}_{\lambda \nu} {h_a}^\lambda\equiv 0,
\end{equation}
where the operator $\nabla^{(\Gamma)}_{\nu}$ is the covariant derivative with respect to the Weitzenb\"{o}ck connection. The connection (\ref{W_connection}) will also be referred to as the canonical connection. The relation (\ref{AP_condition}) is known in the literature as the AP-condition.

The non-commutation of an arbitrary vector fields $V_{a}$ is given by
$$\nabla^{(\Gamma)}_{\nu}\nabla^{(\Gamma)}_{\mu}V_{a}{^{\alpha}} - \nabla^{(\Gamma)}_{\mu}\nabla^{(\Gamma)}_{\nu}V_{a}{^{\alpha}} = R^{\alpha}{_{\epsilon\mu\nu}}
V_{a}{^{\epsilon}} + T^{\epsilon}{_{\nu\mu}} \nabla^{(\Gamma)}_{\epsilon} V_{a}{^{\alpha}},$$
where $R^{\alpha}{_{\epsilon\mu\nu}}$ and $T^{\epsilon}{_{\nu\mu}}$ are the curvature and the torsion tensors of the canonical connection, respectively. The AP-condition (\ref{AP_condition}) together with the above non-commutation formula force the curvature tensor $R^{\alpha}_{~~\mu\nu\sigma}$ of the canonical connection $\Gamma^{\alpha}_{~\mu\nu}$ to vanish identically \cite{NA2007}. Moreover, the parallelization vector fields define a metric tensor on $M$ by
\begin{equation}\label{metric}
g_{\mu \nu} := \eta_{ab}h^{a}{_{\mu}}h^{b}{_{\nu}}
\end{equation}
with inverse metric
\begin{equation}\label{inverse}
g^{\mu \nu} = \eta^{ab}h_{a}{^{\mu}}h_{b}{^{\nu}}.
\end{equation}
The Levi-Civita connection associated with $g_{\mu\nu}$ is
\begin{equation}\label{Christoffel}
\overcirc{\Gamma}{^{\alpha}}{_{\mu\nu}}= \frac{1}{2} g^{\alpha \sigma}\left(\partial_{\nu}g_{\mu \sigma}+\partial_{\mu}g_{\nu \sigma}-\partial_{\sigma}g_{\mu \nu}\right).
\end{equation}
In view of (\ref{AP_condition}), the canonical connection $\Gamma{^\alpha}{_{\mu\nu}}$ (\ref{W_connection}) is metric:
$$\nabla^{(\Gamma)}_{\sigma}g_{\mu\nu}\equiv 0.$$
The torsion tensor of the canonical connection (\ref{W_connection}) is defined as
\begin{equation}
T^\alpha{_{\mu\nu}}:={\Gamma^\alpha}_{\nu\mu}-{\Gamma^\alpha}_{\mu\nu}={h_a}^\alpha\left(\partial_\mu{h^a}_\nu
-\partial_\nu{h^a}_\mu\right).\label{Torsion}\\
\end{equation}
The contortion tensor $K^{\alpha}_{~\mu\nu}$ is defined by
\begin{equation}
K^{\alpha}{_{\mu\nu}} := \Gamma^{\alpha}_{~\mu\nu} - \overcirc{\Gamma}{^{\alpha}}_{\mu\nu}=h_{a}{^{\alpha}}~ \nabla^{(\overcirc{\Gamma})}_{\nu}h^{a}{_{\mu}}. \label{contortion}
\end{equation}
where the covariant derivative $\nabla^{(\overcirc{\Gamma})}_{\sigma}$ is with respect to the Levi-Civita connection. Since $\overcirc{\Gamma}{^{\alpha}}{_{\mu\nu}}$ is symmetric, it
follows that (using (\ref{contortion}))
\begin{equation}\label{torsion-contortion}
T^{\alpha}{_{\mu\nu}} = K^{\alpha}{_{\mu\nu}} - K^{\alpha}{_{\nu\mu}}.
\end{equation}
One can also show that:
\begin{equation}\label{torsion03}
        T_{\alpha \mu \nu}=K_{\alpha \mu \nu}-K_{\alpha \nu \mu},
\end{equation}
\begin{equation}\label{contortion03}
        K_{\alpha \mu \nu}=\frac{1}{2}\left(T_{\nu\alpha\mu}+T_{\alpha\mu\nu}-T_{\mu\alpha\nu}\right),
\end{equation}
where $T_{\mu\nu\sigma} =
g_{\epsilon\mu}\,T^{\epsilon}_{~\nu\sigma}$\, and \,$K_{\mu\nu\sigma} =
g_{\epsilon\mu}\,K^{\epsilon}_{~\nu\sigma}$. It is to be noted that $T_{\mu\nu\sigma}$ is skew-symmetric in the last pair of
indices whereas $K_{\mu\nu\sigma}$ is skew-symmetric in the first pair of indices. Moreover, it follows from (\ref{torsion03}) and (\ref{contortion03}) that the torsion tensor vanishes if and only if the contortion tensor vanishes.
%%%%%%%%%%%%%%%%%%%%%%%%%%%%%%%%%%% Section 2.2 %%%%%%%%%%%%%%%%%%%%%%%%%%%%%%%%%%%%%%%%
\subsection{$f(T)$ gravity}\label{Sec2.2}
%%%%%%%%%%%%%%%%%%%%%%%%%%%%%%%%%%%%%%%%%%%%%%%%%%%%%%%%%%%%%%%%%%%%%%%%%%%%%%%%%%%%%%
The Weitzenb\"{o}ck space is characterized by auto parallelism or absolute parallelism condition, i.e. the vanishing of the tetrad's covariant derivative $\nabla^{(\Gamma)}_{\nu}h^{a}{_{\mu}}\equiv 0$. The derivative operator $\nabla^{(\Gamma)}_{\nu}$ is lacking covariance under local Lorentz transformations (LLT). As a result, all LLT invariant geometrical quantities are allowed to rotate freely in every point of the space \cite{M2013}. Consequently, we cannot fix 16 field variables of the tetrad fields by 10 field variables of the symmetric metric, the extra six degrees of freedom of the 16 tetrad fields need to be fixed in order to identify exactly one physical frame.

In the teleparallel space, there are three independent invariants, under diffeomorphism, may be defined as $I_{1}=T^{\alpha\mu\nu}T_{\alpha\mu\nu}$, $I_{2}=T^{\alpha\mu\nu}T_{\mu\alpha\nu}$ and $I_{3}=T^{\alpha}T_{\alpha}$, where $T^{\alpha}=T_{\nu}{^{\alpha\nu}}$. These can be combined to define the invariant $T=AI_{1}+BI_{2}+CI_{3}$, where $A$, $B$ and $C$ are arbitrary constants \cite{M2013}. For a fixed values of the constants $A=1/4$, $B=1/2$ and $C=-1$ the invariant $T$ is coincide with the Ricci scalar $R^{(\overcirc{\Gamma})}$, up to a divergence term; then a teleparallel version of gravity equivalent to GR can be achieved. The invariant $T$ or the teleparallel torsion scalar is given in the compact form
\begin{eqnarray}
T &:=& {T^\alpha}_{\mu \nu}{S_\alpha}^{\mu \nu},\label{Tor_sc}\\
{S_\alpha}^{\mu \nu}&:=& \frac{1}{2}\left({K^{\mu\nu}}_\alpha+\delta^\mu_\alpha{T^{\beta\nu}}_\beta-\delta^\nu_\alpha{T^{\beta \mu}}_\beta\right),\label{superpotential}
\end{eqnarray}
where the superpotential ${S_\alpha}^{\mu \nu}$ is skew symmetric in the last pair of indices. We next highlight the following useful relation which relates some geometric quantities of physical interests in the Riemannian and the teleparallel geometries.
\begin{equation}\label{B_identity}
\nonumber    R^{(\overcirc{\Gamma})}=-T^{(\Gamma)}-2\nabla^{(\overcirc{\Gamma})}_{\alpha}T^{\nu \alpha}{_{\nu}}.
\end{equation}
Since the second term in the right hand side is a total derivative, the variation of the right (left) hand side with respect to the tetrad (metric) afford the same set of field equations. Using the teleparallel torsion scalar instead of the Ricci scalar in the Einstein-Hilbert action provides TEGR. In spite of this quantitative equivalence they are qualitatively different. For example, the Ricci scalar is invariant under local Lorentz transformation, while the total derivative term is not. Consequently, the torsion scalar is not invariant as well. In conclusion, one can say that the TEGR and GR are equivalent at the level of the field equations, however, at their \textit{lagrangian} level they are not \cite{1010.1041,1012.4039}, a recent modification by considering non-trivial spin connections may solve the problem \cite{Martin2}.
An interesting variant on generalizations of TEGR are the $F(T)$ theories. Similar to the $F(R)$ extensions of Einstein-Hilbert action, one can take the action of $F(T)$ theory as
\begin{equation}\label{action}
{\cal S}({h^a}_\mu, \Phi_A)=\int |h|\left[\frac{M_{\textmd{\tiny Pl}}^2}{2}F(T)+{\cal L}_{m}(\Phi_A)\right]~d^{4}x,
\end{equation}
where $\mathcal{L}_{m}$ is the lagrangian of the matter fields $\Phi_{A}$ and $M_{\textmd{\tiny Pl}}$ is the reduced Planck mass, which is related to the gravitational constant $G$ by $M_{\textmd{\tiny Pl}}=\sqrt{\hbar c/8\pi G}$. Assume the units in which $G = c = \hbar = 1$. In the above equation, $|h|=\sqrt{-g}=\det\left({h^a}_\mu\right)$. For convenience, we rewrite $F(T)=T+f(T)$ and note that the action (\ref{action}) reduces to GR in the case of a vanishing $f(T)$, i.e. $F(T)$ becomes TEGR. The variation of (\ref{action}) with respect to the tetrad field ${h^a}_\mu$ requires the following field equations \cite{BF09,HN15}
\begin{eqnarray}\label{field_eqns}
\nonumber &\left[h^{-1}{h^{a}}_\mu\partial_\rho\left(h{h_{a}}^\alpha
{S_\alpha}^{\rho \nu}\right)-{T^\alpha}_{\lambda \mu}{S_\alpha}^{\nu \lambda}\right]\left(1+f_T\right)\\
&+{S_\mu}^{\rho \nu} \partial_{\rho} T f_{TT}-\frac{1}{4}\delta^\nu_\mu \left(T+f\right)=-4\pi{{\Theta}_{\mu}}^{\nu},
\end{eqnarray}
where $f = f(T)$, $f_{T}=\frac{\partial f(T)}{\partial T}$, $f_{TT}=\frac{\partial^2 f(T)}{\partial T^2}$ and ${{\Theta}_{\mu}}^{\nu}$ is the usual energy-momentum tensor of matter fields. It has been shown that TEGR and GR theories having an equivalent set of field equations. However, their extensions $f(T)$ and $f(R)$, respectively, are not equivalent even at the level of the field equations. The presence of a total derivative term in TEGR \textit{action} would not be reflected in the field equations, so it does not worth to worry about it. However, its presence is crucial when the $f(T)$ extension is considered \cite{1010.1041,1012.4039,Martin2}. As a result, $f(T)$ theories lack the local Lorentz symmetry. Also, it is well known that $f(R)$ theories are conformally equivalent to Einstein-Hilbert action plus a scalar field. In contrast, the $f(T)$ theories cannot be conformally equivalent to TEGR plus a scalar field \cite{Y2011}. Short period these pioneering studies have been followed by a large number of works exploring different aspects of the $f(T)$ gravity in astrophysics \cite{CCDDS11,FF011,FF11,IS12,CGS13,Nashed1,Nashed2,RHTMM13,Nashed3,BFG15,Nashed4,Nashed5,Hanafy:2016} and in cosmology \cite{1205.3421,Nashed:2011,Momeni:2014a,Momeni:2014b,BNO14,BO14,JMM14,HLOS14,NH14,WH14,1503.05281,1503.07427,Waleed:2016,Nunes:2016,BambaOS:2016,OtaloraS:2016}. Some applications show interesting results, e.g. avoiding the big bang singularity by presenting a bouncing solution \cite{CCDDS11,CQSW14,Haro:2014,Bamba:2016}. For more details of $f(T)$-gravity, see the recent review \cite{Saridakis1}.
%%%%%%%%%%%%%%%%%%%%%%%%%%%%%%%%%%%%%%%%%%%%%%%%%%%%%%%%%%%%%%%%%%%%%%%%%%%
\subsection{Modified Friedmann equations}
%%%%%%%%%%%%%%%%%%%%%%%%%%%%%%%%%%%%%%%%%%%%%%%%%%%%%%%%%%%%%%%%%%%%%%%%%%%
In cosmological applications the universe is taken as homogeneous and isotropic in space, i.e. Friedmann-Robertson-Walker (FRW) model, which can be described by the tetrad fields \cite{R32}. It can be written in spherical polar coordinate $x^{0}\equiv t$, $x^{1}\equiv r$, $x^{2}\equiv \theta$ and $x^{3}\equiv \phi$ as follows:
\begin{eqnarray}\label{tetrad}
\nonumber \left({h_{i}}^{\mu}\right)=\hspace{7.5cm}&\\
\nonumber \left(
\begin{tiny}  \begin{array}{cccc}
    1 & 0 & 0 & 0 \\
    0&\frac{L_1 \sin{\theta} \cos{\phi}}{4a(t)} & \frac{L_2 \cos{\theta} \cos{\phi}-4r\sqrt{k}\sin{\phi}}{4 r a(t)} & -\frac{L_2 \sin{\phi}+4 r \sqrt{k} \cos{\theta} \cos{\phi}}{4 r a(t)\sin{\theta}} \\[5pt]
    0&\frac{L_1 \sin{\theta} \sin{\phi}}{4 a(t)} & \frac{L_2 \cos{\theta} \sin{\phi}+4 r \sqrt{k}\cos{\phi}}{4 r a(t)} & \frac{L_2 \cos{\phi}-4 r \sqrt{k} \cos{\theta} \sin{\phi}}{4 r a(t)\sin{\theta}} \\[5pt]
    0&\frac{L_1 \cos{\theta}}{4 a(t)} & \frac{-L_2 \sin{\theta}}{4 r a(t)} & \frac{\sqrt{k}}{a(t)} \\[5pt]
  \end{array}\end{tiny}
\right),&\\
\end{eqnarray}
where $a(t)$ is the scale factor, $L_1=4+k r^{2}$ and $L_2=4-k r^{2}$. The tetrad (\ref{tetrad}) has the same metric as FRW metric
$$ds^2=dt^{2}-a^{2}(t) \left[\frac{dr^{2}}{1- k r^{2}}+r^{2} d\theta^2+r^{2}\sin^{2}(\theta) d\phi^{2}\right].$$
Substituting from the vierbein (\ref{tetrad}) into (\ref{Tor_sc}), we get the torsion scalar
\begin{eqnarray}\label{Tscalar1}
\nonumber   T&=&\frac{6 k- 6 \dot{a}^2}{a^{2}},\\
\nonumber    &=&-6\left(H^2-\frac{k}{a^{2}}\right),\\
             &=&-6H^2(1+\Omega_{k}),
\end{eqnarray}
where the dot denotes the derivative with respect to time $t$, the \textit{Hubble} parameter $H$ is defined as
\begin{equation}\label{Hubble}
H:=\frac{\dot{a}}{a}.
\end{equation}
And the \textit{curvature} energy density parameter $\Omega_{k}$ is defined as
\begin{equation}\label{curv_dens_par}
\Omega_{k}:=\frac{-k}{a^2 H^2}.
\end{equation}
Assume that the material-energy tensor is taken for a perfect fluid $\Theta_{\mu}{^{\nu}}=\textmd{diag}(\rho_{\textmd{m}},-p_{\textmd{m}},-p_{\textmd{m}},-p_{\textmd{m}})$. Using the field equations (\ref{field_eqns}), the modified Friedmann equations for the $f(T)$-gravity in a non-flat FRW background can be written as \cite{BF09,L10,1011.0508}
\begin{eqnarray}
3H^2&=&8\pi\left(\rho_{\textmd{m}}+\rho_{\textmd{eff}}\right)-3\frac{k}{a^{2}},\label{Friedmann1}\\
3(\dot{H}+H^2)&=&-4\pi\left[\rho_{\textmd{m}}+\rho_{\textmd{eff}}+3\left(p_{\textmd{m}}+p_{\textmd{eff}}\right)\right],\label{Friedmann2}
\end{eqnarray}
where the torsion gravity contributes in the field equations as an effective dark sector as
\begin{eqnarray}
\rho_{\textmd{eff}}&=&-\frac{1}{16 \pi}(f+12 H^2 f_{T}),\label{dens1}\\
\nonumber p_{\textmd{eff}}&=&\frac{1}{16 \pi}\left[(f+12 H^2 f_{T}) + 4\dot{H}(f_{T}-12 H^2 f_{TT})\right.\\
&-&\left.\frac{4k}{a^2}(f_{T}+12H^2 f_{TT})\right].\label{press1}
\end{eqnarray}
The TEGR case is restored when $f(T)$ vanishes. One can see that the conservation (continuity) equations when matter and gravity are minimally coupled are
\begin{eqnarray}
% \nonumber to remove numbering (before each equation)
  \dot{\rho}_{\textmd{m}}+3H\left(1+\omega_{\textmd{m}}\right)\rho_{\textmd{m}} &=&0, \\
  \dot{\rho}_{\textmd{eff}}+3H\left(1+\omega_{\textmd{eff}}\right)\rho_{\textmd{eff}} &=&0,\label{cont_eff}
\end{eqnarray}
where the matter EoS parameter $\omega_{\textmd{m}}:=p_{\textmd{m}}/\rho_{\textmd{m}}$ is taken as $\omega_{\textmd{m}}=1/3$ for ultra-relativistic matter (e.g. radiation) or as $\omega_{\textmd{m}}=0$ for non-relativistic matter (e.g. cold matter), while the EoS of the effective torsion gravity $\omega_{\textmd{eff}}:=p_{\textmd{eff}}/\rho_{\textmd{eff}}$ can be identified by (\ref{dens1}) and (\ref{press1}) as
\begin{equation}\label{EoS}
    \omega_{\textmd{eff}}=-1+\frac{4k(f_{T}+12 H^2 f_{TT})}{a^{2}(f+12 H^2 f_{T})}-\frac{4\dot{H}(f_{T}-12H^2 f_{TT})}{(f+12H^2 f_{T})}.
\end{equation}
%+++++++++++++++++++++++++++++++++++++++++++++++++++++++++++++++++++++++++++++++++++++
\subsection{The hidden flat-like universe summary}\label{Sec2.3}
%+++++++++++++++++++++++++++++++++++++++++++++++++++++++++++++++++++++++++++++++++++++
In a recent work we have proposed a hidden class of $f(T)$ gravity constrained by FLU assumptions \cite{HN15}. In the spatially flat universe (SFU) we have shown that there is a good chance to hunt a cosmological constant-like \emph{dark energy} by taking $f_{T}-12H^2f_{TT}=0$, where $T=-6H^{2}$ in the SFU model. This identifies a particular class of $f(T)\propto \sqrt{T}$. However, we have taken a different path allowing evolution away from the cosmological constant without assuming spatial flatness, but enforcing the evolution to be a flat-like. This has been achieved by taking two assumptions \cite{HN15}:
\begin{itemize}
\item [(i)] The first assumption is by taking a vanishing coefficient of $k$ in (\ref{press1}), so
\begin{equation}\label{assump1}
    a^2 f_{T}+12 \dot{a}^2 f_{TT}=0.
\end{equation}
As $f(T)$ in FRW spacetime is a function of time $f(T \rightarrow t)$, one can use the chain rule
\begin{equation}
% \nonumber to remove numbering (before each equation)
  f_{T} = \dot{f}/\dot{T},~~  f_{TT} = \left(\dot{T} \ddot{f}-\ddot{T} \dot{f}\right)/\dot{T}^3.\label{Fd2T}
\end{equation}
Substituting from (\ref{Fd2T}) into (\ref{assump1}), then solve  to $f(T \rightarrow t)$ we get:
\begin{eqnarray}\label{fT1}
\nonumber f(t)&=&\Lambda\\
\nonumber &+&\lambda\int {e^{{\displaystyle{\int}} {\frac{k^2+(3\ddot{a}a-5\dot{a}^2)k+2\ddot{a}^2a^2+4\dot{a}^4-7\dot{a}^2\ddot{a}a+\dot{a}\dddot{a}a^2}
{\dot{a}a(\ddot{a}a-\dot{a}^2+k)}} dt}}dt,\\
\end{eqnarray}
where $\Lambda$ and $\lambda$ are two arbitrary constants.
\item [(ii)] The second is by taking a vanishing coefficient of $k$  in (\ref{fT1}), so
\begin{equation}\label{assump2}
3\ddot{a} a - 5 \dot{a}^2=0.
\end{equation}
The second assumption is clearly independent of the choice of the spatial curvature $k$, while the $f(T)$ itself depends on the choice $k$ as it should be. Otherwise, the three world models, $k=0,\pm 1$, will coincide on each other.
\end{itemize}
Solving (\ref{assump2}) for the scale factor, we get \cite{HN15}
\begin{equation}\label{scale_factor}
    a(t)=a_{0}\left[\frac{3}{3-2H_{0}(t-t_{0})}\right]^{3/2},
\end{equation}
where $a_{0}:=a(t_{0})$, $H_{0}:=H(t_{0})$ and $t_{0}$ is the cutoff time usually taken at Planck's time. In fact the above scale factor and similar ones, e.g. \cite{HLOS14}, can be obtained as a subclass of a more generalized family of scale factors, called $\tilde{q}$-de Sitter \cite{Setare:2016}, where $\tilde{q}$ is a free parameter. This scale factor is an intermediate form between power-law and de Sitter, where both dark energy and dark matter simultaneously are described by this family of solutions. Then the Hubble parameter will be
\begin{equation}\label{Hubble}
    H=\frac{3H_{0}}{3-2H_{0}(t-t_{0})}.
\end{equation}
Also, the deceleration parameter $q:=-\frac{a\ddot{a}}{\dot{a}^{2}}=-5/3$. At a large-Hubble regime, the density can be given as
$$\rho_{\textmd{eff}} \rightarrow \frac{1}{16 \pi}\left[\Lambda+\frac{81 a_{0}^{2}H_{0}^{2}\lambda}{(3+2H_{0} t_{0})^{3}}\right],$$
which agrees with vacuum energy predictions of cosmic inflation. Finally, we see that the FLU assumption (ii) gives a scale factor consistent with cosmic inflation. This makes the choice $t \thickapprox 0$ is a natural choice, when dealing with the FLU model. Substituting from (\ref{scale_factor}) into (\ref{assump1}), taking the power series at $t=0$ we get \cite{HN15}
\begin{equation}\label{f(t)}
    f(t)=\sum_{n=0}^{\infty}c_{n}t^{n},
\end{equation}
where $n$ is a positive integer and the coefficients $c_{n}$ are evaluated in terms of $\{t_{0},a_{0},H_{0},k,\lambda,\Lambda\}$, e.g. $c_0 = \Lambda$, $c_1=-2\tau_0 H_0 \lambda k - \frac{81a_0^2 \lambda}{8\tau_0^4 H_0^2}$, $c_2=-\frac{81a_0^2\lambda}{4\tau_0^5 H_0^2}-\frac{16\tau_0^5 H_0^4 \lambda k^2}{81a_0^2}$, ... etc, where $\tau_0=t_{0}+\frac{3}{2H_0}$.
Substituting from (\ref{scale_factor}) into (\ref{Tscalar1}) the torsion scalar can be given as
\begin{equation}\label{Tsc}
    T(t)=\frac{6 k [3-2H_{0}(t-t_{0})]^{5}-1458 a_{0}^{2}H_{0}^{2}}{27 a_{0}^{2} [3-2H_{0}(t-t_{0})]^{2}}.
\end{equation}
The inverse relation of (\ref{Tsc}) enables us to express the time as a function of the teleparallel torsion scalar. After, easy arrangements we can get the conventional form of the FLU $f(T)$ theory as
\begin{equation}\label{fT}
f(T)\propto\sum_{n=0}^{\infty}\frac{\alpha_{n}}{\sqrt{(-T)^{n}}},
\end{equation}
where the coefficients $\alpha_{n}$ consist of the same constants as the coefficients $c_{n}$. At inflation stage, where $T \rightarrow \infty$, the above $f(T)$ reduces to a constant, i.e. $f(T)=\alpha_{0}=\Lambda$, then the effective torsion gravity acts just like a cosmological constant with EoS $\rho_{\textmd{eff}}=-p_{\textmd{eff}}$. When $T$ decreases, deviations are strongly expected, which again supports the choice of the torsion gravity in the FLU model to describe inflationary universe epoch.\\

To summarize the FLU model, we note that, in $f(T)$ theories one needs two assumptions to obtain an $f(T)$ theory. This can be done by imposing a particular choice of the scale factor in addition to a fixed EoS parameter as inputs. As a matter of fact, the matter contribution is usually neglected during inflation. In addition, choosing a fixed EoS of the effective density and pressure of the torsion gravity during inflation is far from being correct, because we cannot assume a particular choice of the EoS during this period. However, it is more convenient to impose a physical constraint consistent with the cosmic inflation leaving the effective EoS to be identified by the model as a dynamical output parameter as usually done in the scalar field inflationary models.\\

In the FLU, we also need two assumptions to constrain the model. This is done by imposing the model assumptions (\ref{assump1}) and (\ref{assump2}) to be consistent with the FLU, aiming to enforce the universe to act as the flat model whatever the choice of $k$. As a result, we have obtained the solution pairs ($a(t)$, $f(T)$) as given by (\ref{scale_factor}) and (\ref{f(t)}), respectively. The model shows a satisfactory agreement with cosmic inflation requirements. Then the model provides a dynamical effective EoS parameter as an output. Consequently, the continuity equation, i.e. Friedmann equations are automatically satisfied. This is similar to the quintessence (scalar field) model, but here it is powered by a torsion gravity model. As mentioned in \cite{HN15}, the FLU $f(T)$ theory does not reduce to TEGR. In other words, the torsion fluid is not an ordinary matter, as it should be during inflation, and cannot be covered by the conventional TEGR or GR. This supports our choice of leaving the EoS parameter to be identified by the model. Finally, we find that the EoS, (\ref{EoS}), for the three models $k=0,\pm 1$, at a time large enough, evolves similarly to same fate jumping by a quantized value $\frac{2}{9}$ such that $$\omega_{\textmd{eff}}=\frac{p_{\textmd{eff}}}{\rho_{\textmd{eff}}}=\left\{-1, -\frac{7}{9}, -\frac{5}{9}, -\frac{1}{3},\cdots\right\},$$
as $n$ increases by unity. So the FLU model has succeeded to enforce the cosmic behaviour to the flat limit by the end of the inflation period.
%%%%%%%%%%%%%%%%%%%%%%%%%%%%%%%%%%%%%%%%%%%%%%%%%%%%%%%%%%%%%%%%%%%%%%%%%%%%
\section{Single field with double slow roll solutions}\label{Sec3}
%%%%%%%%%%%%%%%%%%%%%%%%%%%%%%%%%%%%%%%%%%%%%%%%%%%%%%%%%%%%%%%%%%%%%%%%%%%%
Assuming that the inflation epoch is dominated by the scalar field potential only. The slow roll models define two parameters as \cite{LL2000}
\begin{equation}\label{slow_roll}
    \epsilon=\frac{1}{16\pi}\left(\frac{V'}{V}\right)^{2},\qquad \eta=\frac{1}{8\pi}\left(\frac{V''}{V}\right).
\end{equation}
These parameters are called slow roll parameters. Consequently, the slow roll inflation is valid where $\epsilon \ll 1$ and $|\eta| \ll 1$ when the potential is dominating, while the end of inflation is characterized by $\textmd{Max}(\epsilon,|\eta|)=1$ as the kinetic term contribution becomes more effective. The slow roll parameters define two observable parameters
\begin{equation}\label{r_n}
r=16\epsilon,\quad n_{s}=1-6\epsilon+2\eta,
\end{equation}
where $r$ and $n_{s}$ are called the tensor-to-scalar ratio and scalar tilt, respectively. Recent observations by Planck and BICEP2 measure almost the same scalar tilt parameters $n_{s}\sim 0.96$. However, Planck puts an upper limit $r<0.11$ which supports models with small $r$, the BICEP2 sets a lower limit on the $r>0.2$ which supports inflationary models with large $r$. We devote this Section to investigate the capability of the slow roll models to perform both Planck and BICEP2.

It is clear that Planck and BICEP2 observations agree on the scalar tilt parameter value $n_{s}\sim 0.963$, while they give different tensor-to-scalar ratios $r$. In order to construct a scalar potential performing Planck and BICEP2 data, we found that if the slow roll parameters (\ref{slow_roll}) satisfy the proportionality relation $\epsilon\propto \eta^{2}$, this gives a chance to find two values of $\eta$ performing the same $n_{s}$ but two different values of $\epsilon$. Consequently, there are two values of $r$. This can be achieved as follows: Using (\ref{r_n}) and the proportionality relation we have
\begin{eqnarray}\label{epseta}
\nonumber n_{s}&=&1-6(\epsilon=\kappa\eta^{2})+2\eta,\\
\textmd{i.e.} \quad \eta^{\pm} &=&\frac{1}{6\kappa}\left(1\pm\sqrt{1+6\kappa(1-n_{s})}\right),\label{quad_eta}
\end{eqnarray}
where $\kappa$ is a constant coefficient and $\eta^{\pm}$ are due to the $\pm$ discriminant. It is clear that there are possibly two different values of the parameter $\eta$ for a single scalar tilt  parameter $n_{s}$. Accordingly, we have $\epsilon=\kappa\eta^{2}=16/r$ which provides double values of $r$ for each $n_{s}$. More concretely, assuming the scalar tilt parameter $n_{s}=0.963$ \cite{Pl2} and substituting into (\ref{epseta}), for a particular choice of the constant $\kappa = 30$; then we calculate two possible values of $\eta$ as
\begin{itemize}
\item [(i)] The first solution $\eta^{+}$ has a positive value of $\sim 2.09\times10^{-2}$ which gives $\epsilon^{+} \sim 1.31\times10^{-2}$.
\item [(ii)] The second solution of $\eta^{-}$ has a negative value of $\sim -9.34\times10^{-3}$ which leads to $\epsilon^{-} \sim 3.05\times10^{-3}$.
\end{itemize}
\begin{figure}[t]
\begin{center}
\includegraphics[scale=0.4]{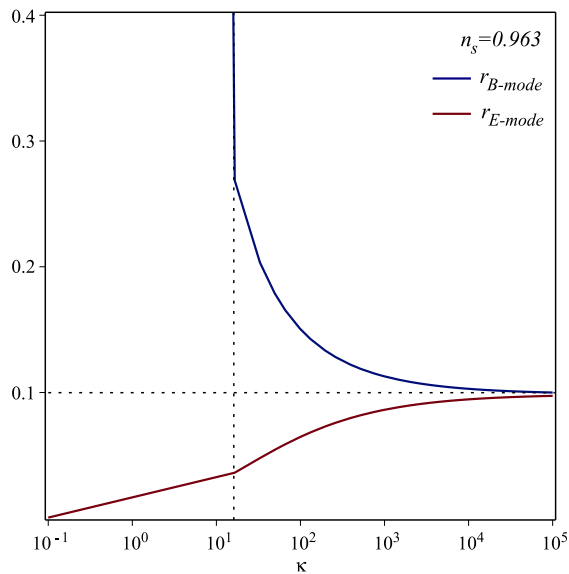}
\caption{At small conformal weight $\kappa$, the model predicts $E$- and $B$-modes of polarization. At large $\kappa$, the model predicts only $E$-mode one.}
\label{kappa}
\end{center}
\end{figure}
Surely both positive and negative values of $\eta$ give the same scalar tilt $n_{s} \sim 0.963$. Nevertheless, we can get two simultaneous tensor-to-scalar ratios: The first is $r^{+} \sim 0.21$, while the other is smaller $r^{-} \sim 4.9\times10^{-2}$. We conclude that the slow roll inflationary models which are characterized by the proportionality $\epsilon \propto \eta^{2}$ can perform both $E$-mode and $B$-mode polarizations \cite{BHS13}, when a negative value of $\eta$ is observed near the peak of $\varphi$, it would need to be offset by a positive value of $\eta$ at some later time over a comparable field range in order to get $\epsilon$ to be small again during the period of observable inflation. Generally, at low values of $\kappa$, the model predicts one small value of $r$ as required by the $E$-mode polarization models, in addition to another higher value as required by the $B$-mode polarization inflationary models, see Figure \ref{kappa}. Interestingly, at large values of $\kappa$, the model predicts a single value of the tensor-to-scalar parameter $r^{\pm}\rightarrow 0.0987$, which agrees with the upper Planck limit $r_{0.002}<0.11$ at $95\%$ CL.

Moreover, we can investigate the potential pattern which is characterized by the proportionality relation $\epsilon=\kappa\eta^{2}$. Recalling (\ref{slow_roll}), this relation provides a simple differential equation
\begin{equation}
\nonumber \kappa~V''^{2}-4\pi~V'^{2}=0,
\end{equation}
with a solution
\begin{equation}\label{Planck_BICEP2}
V(\varphi)=A+Be^{\pm2\sqrt{\frac{\pi}{\kappa}}\varphi},
\end{equation}
where $A$ and $B$ are constants of integration. In this way, we found that Starobinsky model might be reconstructed naturally from observations if we want our model to perform $E$-mode and $B$-mode polarizations.
%%%%%%%%%%%%%%%%%%%%%%%%%%%%%%%%%%%%%%%%%%%%%%%%%%%%%%%%%%%%%%%%%%%%%%%%%%%%
\section{Gravitational quintessence models}\label{Sec4}
%%%%%%%%%%%%%%%%%%%%%%%%%%%%%%%%%%%%%%%%%%%%%%%%%%%%%%%%%%%%%%%%%%%%%%%%%%%%
As we mentioned in the introduction of this work, due to the lack of the conformal invariance of the $f(T)$ theories we have developed an alternative technique to map the torsion contribution in the modified Friedmann equations into a quintessence scalar field. This technique could allow to induce a gravitational quintessence model from an $f(T)$ model, or inversely to reconstruct $f(T)$ gravity from a quintessence potential. In what follows we give a brief of the used method
\subsection{Torsion potential}
In the FLU model, we have assumed the case when the torsion potential is constructed by a scalar field $\varphi$. This consideration suggests that the torsion and the contortion to have the following semi-symmetric forms \cite{HN15}
\begin{eqnarray}
    T^{\alpha}{_{\mu\nu}}&=&\sqrt{3/2}\left(\delta^{\alpha}_{\nu}\partial_\mu \varphi-\delta^{\alpha}_{\mu}\partial_\nu \varphi\right),\label{semi-symm-torsion}\\
    K^{\mu\nu}{_{\alpha}}&=&\sqrt{3/2}\left(\delta^{\nu}_{\alpha}\partial^\mu \varphi-\delta^{\mu}_{\alpha}\partial^\mu \varphi\right),\label{semi-symm-contortion}
\end{eqnarray}
where $\partial^{\mu}=g^{\mu\nu}\partial_{\nu}$. Substituting from (\ref{semi-symm-torsion}) and (\ref{semi-symm-contortion}) into (\ref{Tor_sc}) the teleparallel torsion scalar can be related to the gradient of the scalar field as
\begin{equation}\label{Tsc_phi}
    T=-9~\partial_\mu\varphi~ \partial^\mu\varphi.
\end{equation}
This relation is the cornerstone of the scalar field model of this section. It allows to define a scalar field induced by the symmetry of the spacetime through the teleparallel torsion scalar. In order to compare this model to the standard inflation models and to simplify the calculations, we take the flat universe case. By combining the relation (\ref{Tsc_phi}) with (\ref{Tsc}); then the kinetic term of the scalar field can be related to the cosmic time by
\begin{equation}\label{kin}
\dot{\varphi}^2=\frac{3}{2(t-\tau_{0})^2},
\end{equation}
where $\tau_0=t_0+\frac{3}{2H_0}$. The integration of (\ref{kin}) with respect to time gives the scalar field as
\begin{equation}\label{phi}
    \varphi=\varphi_{0} \pm \sqrt{6}/2\ln{(t-\tau_{0})}.
\end{equation}
In order to investigate the relation between the teleparallel torsion scalar and the inflaton field we perform the following comparisons. At strong coupling the inflaton can be related to the canonical scalar field $\Omega$ as \cite{KLR214}
\begin{equation}\label{canonical}
\varphi=\pm \sqrt{\frac{3}{2}}\log{\Omega},
\end{equation}
As a result, we have shown that the teleparallel torsion scalar
\begin{equation}\label{canon}
    T(\varphi)=-\frac{27}{2} e^{\pm 2\sqrt{2/3}(\varphi-\varphi_{0})},
\end{equation}
plays the role of the canonical scalar field. Although introducing an inflationary phase at an early universe stage leads to solve some problems of standard cosmology, it has not yet made direct connection with a unique fundamental theory. As a consequence, there is no way to justify the main features of an inflationary model (e.g. the shape of the potential). In our investigation of the FLU model \cite{HN15}, we have shown that a quintessence potential can be induced by $f(T)$ gravity. The mapping from the teleparallel torsion to the scalar field $\varphi$, (\ref{canon}), allows to write the effective density and the pressure as
\begin{equation}\label{press_phi1}
    \rho_{\textmd{eff}}\rightarrow \rho_{\varphi}=\frac{1}{2}\dot{\varphi}^2+V(\varphi),\quad p_{\textmd{eff}}\rightarrow p_{\varphi}=\frac{1}{2}\dot{\varphi}^2-V(\varphi).
\end{equation}
In absence of matter, the Friedmann equation (\ref{Friedmann1}) reads
\begin{equation}\label{Hubble_sc}
\nonumber    H^2=\frac{8\pi}{3}\left(\frac{1}{2}\dot{\varphi}^2+V(\varphi)\right)-\frac{k}{a^{2}},
\end{equation}
and the continuity equation (\ref{cont_eff}) transforms to the Klein-Gordon equation
\begin{equation}
\nonumber    \ddot{\varphi}+3H\dot{\varphi}+V'(\varphi)=0.
\end{equation}
\begin{figure}
\centering
\includegraphics[scale=.3]{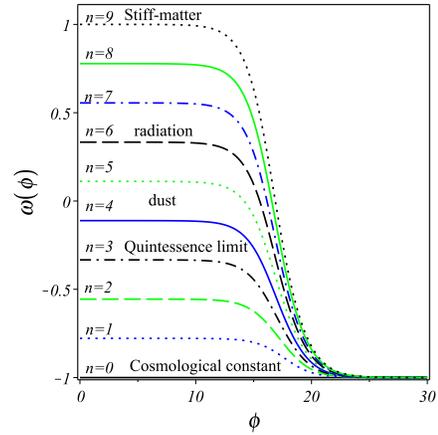}
\caption{The evolution of the scalar field EoS-parameter. The initial conditions of the set of parameters $\{t_{0},a_{0},H_{0},\Lambda\}$ are taken to fit with an early time phase as $a_{0}=10^{-9}$, $H_{0}=10^{9}~s^{-1}$, $t_{0}=t_{\textmd{\tiny Pl}}=10^{-44}~s$, $\varphi_{0}=2.7$, $\lambda=0.3$ and $\Lambda=10^{-30}~s^{-2}$.}
\label{Fig2}
\end{figure}
Substituting from (\ref{phi}) into (\ref{press1}), we write the proper pressure in terms the scalar field as
\begin{equation}\label{dens_phi}
p_{\varphi}=-\frac{\Lambda }{16 \pi}-\lambda \sum_{n=0} \beta_{n} e^{-{n\sqrt{2/3}}(\varphi-\varphi_{0})}.
\end{equation}
Same way can be used to identify $\rho_{\varphi}$. As a result and to make sure that the torsion contribution has been transformed into a scalar field. In Figure \ref{Fig2}, we plot the evolution of the EoS-parameter $\omega_{\varphi}=p_{\varphi}/\rho_{\varphi}$ of the obtained scalar field for different values of $n$. The plots of Figure \ref{Fig2} show clearly that the EoS of the scalar field goes from a ground value at $\omega_{\varphi}=-1$ to higher values by a quantized values $2/9$ as the scalar field decays at different values of $n$. This result matches perfectly our previous results using the classical treatment of $f(T)$ field equations \cite{HN15}. In particular for $n=3$, the EoS evolves as $\omega_{\varphi}: -1 \rightarrow -1/3$, which is in agreement with requirements of the cosmic inflation phase. We have suggested that the models with $n<3$ to represent cosmic inflation phases, while the $n>3$ models can be used for a later stage as $\omega>-1/3$ which is more suitable for kinetic dominant stages after inflation. In addition, we see that there is no physical motivations to study models with $n>9$ as $\omega_{\varphi}>1$ which does not represent a physical matter so far. Finally, we can identify the induced potential as \cite{HN15}
\begin{equation}\label{sc_pot}
V(\varphi)=\frac{\Lambda }{16\pi}+\frac{3}{4} e^{2\sqrt{2/3}(\varphi-\varphi_{0})}+\lambda \sum_{n=0} \beta_{n} e^{-n\sqrt{2/3}(\varphi-\varphi_{0})}.
\end{equation}

\begin{figure}
\centering
\subfigure[$~n=0$]{\label{Fig3a}\includegraphics[scale=.22]{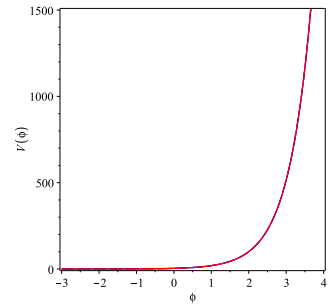}}
\subfigure[$~n=1$]{\label{Fig3b}\includegraphics[scale=.22]{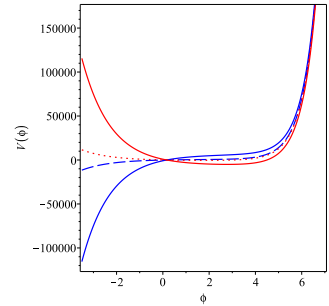}}
\subfigure[$~n=2$]{\label{Fig3c}\includegraphics[scale=.22]{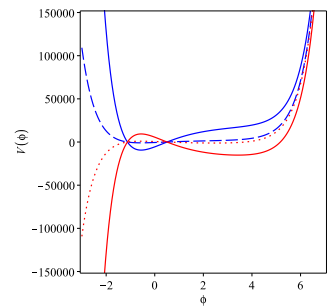}}
\caption[figtopcap]{Schematic plots of the scalar field potentials according to the order of expansion of the $f(T)$ function. The initial conditions of the set of parameters $\{t_{0},a_{0},H_{0},\Lambda\}$ are taken to fit with an early time phase as $a_{0}=10^{-9}$, $H_{0}=10^{9}~s^{-1}$, $t_{0}=t_{\textmd{\tiny Pl}}=10^{-44}~s$ and $\Lambda=10^{-30}~s^{-2}$. The blue solid lines are for $\lambda>1$, the blue dash lines are for $0<\lambda<1$, the red dot lines are for $0>\lambda>-1$ and the red solid lines are for $\lambda<-1$.}\label{Fig3}
\end{figure}

It is worth to mention that the obtained potential covers different classes of inflation. For example, the ($n=0$)-model produces the following potential
$$V_{0}=\frac{\Lambda }{16\pi}+\frac{3}{4} e^{2\sqrt{2/3}(\varphi-\varphi_{0})},$$
which is a combination between the cosmological constant density (the first term) and the kinetic term (the second term). When the kinetic term is comparable to the cosmological constant density, the ($n=0$)-model is typical to that obtained by (\ref{Planck_BICEP2}). In conclusion, the potential pattern can perform both $E$ and $B$ polarization modes \cite{HN15}. However, the potential never drops to zero and the inflation never ends so that it needs an extra mechanism to end the inflation. If the kinetic term is negligible the model is a typical de-Sitter which can be considered as a useful model in the late cosmic acceleration rather than in the early cosmic acceleration phases. Also, the ($n=1$)-model produces the following potential
\begin{eqnarray}\label{V1}
\nonumber V_{1}&=&\frac{\Lambda }{16\pi}-\frac{81 \lambda a_{0}^{2}}{128 \pi H_{0}^{2}\tau_{0}^{3}}+\frac{3}{4} e^{2\sqrt{2/3}(\varphi-\varphi_{0})}\\
&-&\frac{63 \lambda a_{0}^{2}}{64 \pi H_{0}^{2}\tau_{0}^{4}}e^{-\sqrt{2/3}(\varphi-\varphi_{0})}.
\end{eqnarray}
The model allows to discuss two possible situations: (i) If the kinetic term is not negligible, the potential shares some features of the potentials required for cyclic universe models. (ii) If the slow roll approximation is considered, i.e. negligible kinetic term, the model gives a quasi power law inflation which will be discussed in details in the next section. Moreover, it has been shown that the ($n=2$)-model produces Starobinsky-like model \cite{HN15}. The patterns of the potential are plotted in Figures \ref{Fig3}\subref{Fig3a}-\ref{Fig3}\subref{Fig3c}.
%+++++++++++++++++++++++++++++++++++++++++++
\subsection{Quasi inverse power law inflation}
%+++++++++++++++++++++++++++++++++++++++++++
In this section we discuss the induced potential up to $n=1$ in more details. In this model the term which contains $\sim e^{\varphi}$ decreases as $e^{-\varphi}$ increases. This tells that the kinetic term can be considered as negligible at certain time which matches the slow roll condition $\dot{\varphi}^{2}\ll V(\varphi)$, then the effective slow roll potential (\ref{V1}) reduces to
\begin{equation}\label{eff_pot}
    V_1=\frac{\Lambda}{16\pi}\left[1-\frac{9\lambda a_{0}^{2}}{8\Lambda \tau_{0}^{4}H_{0}^{2}}\left(9\tau_{0}+14e^{-\sqrt{2/3}(\varphi-\varphi_{0})}\right)\right].
\end{equation}
It is clear that the effective potential is powered by an exponential function in the scalar field. Inflation with an exponential potential is also called power law inflation, because this type of inflation is characterized by a scale factor $a(t) \propto t^{p}$, where $p>1$ \cite{LM1985}. The exponential potential of the power law inflation model never drops to zero. As a result, this model needs an extra mechanism to end the inflation and can be considered as an incomplete model. Fortunately, the potential (\ref{eff_pot}) drops to zero as
$$\frac{9\lambda a_{0}^{2}}{8\Lambda \tau_{0}^{4}H_{0}^{2}}\left(9\tau_{0}+14e^{-\sqrt{2/3}(\varphi-\varphi_{0})}\right) \sim 1.$$ This allows inflation to end naturally with no need to an extra mechanism. We calculate the slow roll parameters defined by (\ref{slow_roll}) of the potential (\ref{eff_pot}), the parameters are
\begin{eqnarray}
\nonumber  \epsilon_{1} &=& \frac{1323 \lambda^{2}a_{0}^{4}e^{-2\sqrt{2/3}(\varphi-\varphi_{0})}}{2\pi\left[8\Lambda\tau_{0}^{4}H_{0}^{2}-9\lambda a_{0}^{2}\left(9\tau_{0}+14e^{-\sqrt{2/3}(\varphi-\varphi_{0})}\right)\right]^{2}},\label{eps}\\ \\
\nonumber  \eta_{1} &=& -\frac{21 \lambda a_{0}^{2}e^{-\sqrt{2/3}(\varphi-\varphi_{0})}}{2\pi\left[8\Lambda\tau_{0}^{4}H_{0}^{2}-9\lambda a_{0}^{2}\left(9\tau_{0}+14e^{-\sqrt{2/3}(\varphi-\varphi_{0})}\right)\right]}.\label{eta}\\
\end{eqnarray}
The subscript $1$ is used to refer to the ($n=1$)-model or simply the potential (\ref{eff_pot}). The slow roll parameters (\ref{eps}) and (\ref{eta}) show that the model is satisfying the proportionality $\epsilon=\kappa \eta^{2}$, where the proportionality constant in this case is $\kappa=6 \pi$. Remarkably, this relation is not independent of the choice of the initial conditions, even it allows the vanishing of the cosmological constant. The number of $e$-folds $N$ from the end of inflation to the time of horizon crossing for observable scales is given by
\begin{eqnarray}\label{efold}
\nonumber    &&N_{*}(\varphi)=-8\pi\int_{\varphi}^{\varphi_{f}}{\frac{V}{V'}}d\varphi\\
\nonumber   &&=4\sqrt{6}\pi\left[\varphi-\frac{2\sqrt{6}\tau_{0}}{63\lambda a_{0}^{2}}\left(\Lambda \tau_{0}^{3}H_{0}^{2}-\frac{81}{8}\lambda a_{0}^{2}\right)e^{\sqrt{2/3}(\varphi-\varphi_{f})}\right],\\
   &&
\end{eqnarray}
where $\varphi_{f}$ is the value of the scalar field at the end of inflation. At the end of inflation, i.e. Max($\epsilon$, $|\eta|$)=1, we have $\varphi_{f}=2.7$. In order to relate the slow roll parameters to the number of $e$-folds before the end of inflation, we use the inverse relation of the above equation\footnote{The asymptotic behavior of LambertW function is defined as \cite{corless1996lambertw}
$$\textmd{\small LambertW}(x)\sim \log(x)-\log[\log(x)]+\sum_{m = 0}^{\infty}\sum_{n = 0}^{\infty}C^{m}_{ n}\frac{\log[\log(x)]^{m+1}}{\log(x)^{m+n+1}}.$$}
\begin{equation}\label{inv_rel}
    \varphi=-\frac{\sqrt{6}}{24\pi}~f(N),
\end{equation}
where
\begin{eqnarray}
\nonumber f(N)&=&\left[N-12\pi\textmd{LambertW}\left(\frac{\tau_{0}\left(81 \lambda a_{0}^{2}-8\Lambda \tau_{0}^{3}H_{0}^{2}\right)}{126 \lambda a_{0}^{2}}\right.\right.\\
&\times& \left.\left.e^{\frac{N-4\sqrt{6}\pi \varphi_{0}}{12\pi}}\right)\right].
\end{eqnarray}
Here LambertW function satisfies $\textmd{LambertW}(x)e^{\textmd{LambertW}(x)}=x$. Now the slow roll parameters of the model read
\begin{equation}\label{epsN}
\epsilon(N)=\frac{1323\lambda^{2} a_{0}^{4}~e^{\frac{-f(N)}{6\pi}}}{2\pi\left[8\Lambda\tau_{0}^{4}H_{0}^{2}-81\lambda a_{0}^{2}\tau_{0}-126\lambda a_{0}^{2}~e^{\frac{-f(N)}{9\pi}}\right]^{2}},
\end{equation}
\smallskip
and
\begin{equation}\label{etaN}
\eta(N)=\frac{-21\lambda a_{0}^{2}~e^{\frac{-f(N)}{9\pi}}}{2\pi\left[8\Lambda\tau_{0}^{4}H_{0}^{2}-81\lambda a_{0}^{2}\tau_{0}-126\lambda a_{0}^{2}~e^{\frac{-f(N)}{9\pi}}\right]}.
\end{equation}
\begin{figure}
\centering
\includegraphics[scale=.4]{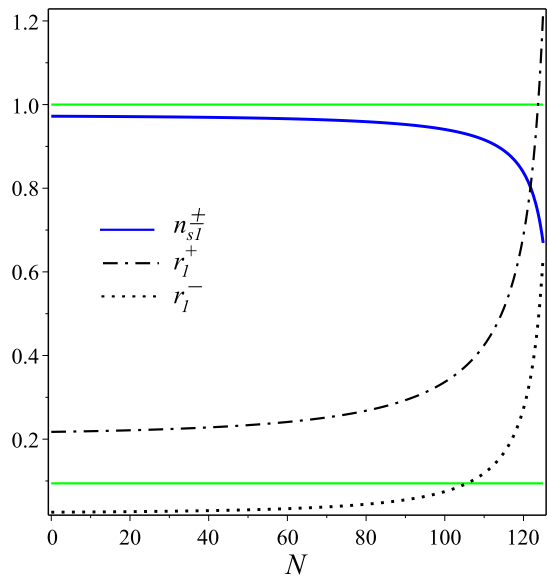}
\caption{The slow roll parameters vs the $e$-folding number of $V_{1}$-model (\ref{eff_pot}). The initial conditions of the set of parameters $\{t_{0},a_{0},H_{0},\Lambda\}$ are taken to fit with an early time phase as $a_{0}=10^{-9}$, $H_{0}=10^{9}~s^{-1}$, $t_{0}=t_{\textmd{\tiny Pl}}=10^{-44}~s$, $\Lambda=10^{-30}~s^{-2}$ and $\lambda=3$.}
\label{Fig4}
\end{figure}
\begin{table}[t]
\caption{The predicted parameters of the $V_{1}$-model (\ref{eff_pot})}
\label{T1}
\begin{tabular*}{\columnwidth}{@{\extracolsep{\fill}}lrrc@{}}
\hline
$N$                & $55\qquad$            & $66\qquad$            & $95\quad$             \\
\hline
$\eta_{1}^{+}$     & $2.80\times 10^{-2}$  & $2.86\times 10^{-2}$  & $3.22\times 10^{-2}$  \\[6pt]
$\eta_{1}^{-}$     & $-1.03\times 10^{-2}$ & $-1.09\times 10^{-2}$ & $-1.45\times 10^{-2}$  \\[6pt]
$\epsilon_{1}^{+}$ & $1.48\times 10^{-2}$  & $1.54\times 10^{-2}$  & $1.95\times 10^{-2}$  \\[6pt]
$\epsilon_{1}^{-}$ & $2.02\times 10^{-3}$  & $2.26\times 10^{-3}$  & $3.95\times 10^{-3}$  \\[6pt]
$n_{s1}^{\pm}$     & $0.967\qquad$         & $0.965\qquad$         & $0.947\quad$           \\[6pt]
$r_{1}^{+}$        & $0.237\qquad$         & $0.247\qquad$         & $0.312\quad$          \\[6pt]
$r_{1}^{-}$        & $3.23\times 10^{-2}$  & $3.61\times 10^{-2}$  & $6.31\times 10^{-2}$  \\
\hline
\end{tabular*}
\end{table}
In the following we show the model capability to predict double tensor-to-scalar ratios $(\epsilon^{+},\epsilon^{-})$ for a single scalar tilt $n_{s}^{\pm}$ value at a chosen suitable $e$-folding in the allowed range $30<N_{*}<60$. For example at $N_{*}=55$, the direct substitution in (\ref{epsN}) and (\ref{etaN}) identifies the values of the slow roll parameters as $\epsilon_{1}(N_{*})=1.48\times 10^{-2}$ and $\eta_{1}(N_{*})=2.80\times 10^{-2}$. As a result, the inflation parameters are evaluated as $n_{s1}= 0.967$ and $r_{1}=0.237$ which gives a tensor-to scalar ratio exceeding the Planck upper limit, while the relation $\epsilon=\kappa \eta^{2}$ provides another hidden values. These values can be obtained by carrying out the following steps: To distinguish the new values from the just obtained ones, we use superscript $+$ for them, i.e. $\eta_{1}^{+}$, $\epsilon_{1}^{+}$, and so on. Using equation (\ref{quad_eta}) we have another hidden value of the slow roll parameter $\eta_{1}$, we denotes its value as $\eta_{1}^{-}$. At $N_{*}=55$, we obtain its value as $\eta_{1}^{-}=-1.03\times 10^{-2}$, it follows by another hidden slow roll parameter $\epsilon_{1}^{-}=\kappa (\eta_{1}^{-})^{2}=2.02\times 10^{-3}$. Remarkably, the last pairs are perfectly match the Planck satellite data \cite{Pl2}; then the second pairs of the inflation parameters are given as follows: The scalar tilt as expected has the same value $n_{1,s}^{-}=n_{s1}^{+}=0.967$, so we denotes the unique value of the scalar tilt of this model as $n_{1,s}^{\pm}=0.976$, while the tensor-to-scalar ratio is $r_{1}^{-}=3.23\times 10^{-2}$. As clear the last inflation parameters' pairs are in agreement with the Planck data. In conclusion, we find that the ($n=1$) slow roll inflation model which is characterized by the relation $\epsilon_{1}=\kappa \eta_{1}^{2}$ can perform a double attractors of the inflation: The first predicts an $B$ mode polarization with a large tensor-to-scalar ratio $r_{1}^{+}=0.237$ in addition to a second solution predicting $E$ mode with a small tensor-to-scalar ration $r_{1}^{-}=3.23\times 10^{-2}$, while both predict a unique value of the scalar tilt $n_{s1}^{\pm}=0.967$. In Figure \ref{Fig4}, we plot the double pairs of the inflation parameters vs the number of $e$-folds before the end of the inflation. Also, we give in Table \ref{T1} a detailed list of the parameters expected by the model at different $e$-folding numbers.
%%%%%%%%%%%%%%%%%%%%%%%%%%%%%%%%%%%%%%%%%%%%%%%%%%%%%%%%%%%%%%%%%%%%%%%%%%%%
\section{Conclusions}\label{Sec5}
%%%%%%%%%%%%%%%%%%%%%%%%%%%%%%%%%%%%%%%%%%%%%%%%%%%%%%%%%%%%%%%%%%%%%%%%%%%%
In the present article, we have summarized the FLU model. In which we have identified a special class of $f(T)$ gravity model, that cannot by covered by TEGR or GR theories. The FLU has been verified to be consistent with the early cosmic inflation.

We have discussed the potential pattern fulfilling the requirements of performing both $E$ and $B$ modes of the CMB polarization. Also, we have studied a new technique to induce a scalar field potential from $f(T)$ gravity by considering that the teleparallel torsion is a semi-symmetric one when the torsion potential is made of a scalar field.

By applying the slow roll conditions, we have shown that the derived potential up to $n=1$ can be classified as a quasi inverse power law inflation model. However, its potential contains an additive constant allowing the potential to drop to zero. This can provide a graceful exit inflation model with no need to an extra mechanism as in the power law inflation model.

We have shown that the obtained potential is coincide to the generic potential which allows $E$ and $B$ modes of polarization. The calculated slow roll parameters can be split into two different solutions allowing double values of the tensor-to-scalar ratio: At $N_{*}=55$, the first is small enough to match the Planck satellite data $r_{1}^{-}=3.23 \times 10^{-2}$, while the second is large enough to produce gravitational waves at the end of inflation as $r_{1}^{+}=0.237$. However, the double solutions predict exactly the same scalar title as $n_{s1}^{\pm}=0.967$.
%%%%%%%%%%%%%%%%%%%%%%%%%%%%%%%%%%%%%%%%%%%%%%%%%%%%%%%%%%%%%%%%%%%%%%%%%%%%%%%%%%%%%%
\subsection*{Acknowledgments}
This article is partially supported by the Egyptian Ministry of Scientific Research under project No. 24-2-12.
%%%%%%%%%%%%%%%%%%%%%%%%%%%%%%%%%%%%%%%%%%%%%%%%%%%%%%%%%%%%%%%%%%%%%%%%%%%%%%%%%%%%%%
\bibliographystyle{apsrev}
%\bibliography{References}

\begin{thebibliography}{69}
\expandafter\ifx\csname natexlab\endcsname\relax\def\natexlab#1{#1}\fi
\expandafter\ifx\csname bibnamefont\endcsname\relax
  \def\bibnamefont#1{#1}\fi
\expandafter\ifx\csname bibfnamefont\endcsname\relax
  \def\bibfnamefont#1{#1}\fi
\expandafter\ifx\csname citenamefont\endcsname\relax
  \def\citenamefont#1{#1}\fi
\expandafter\ifx\csname url\endcsname\relax
  \def\url#1{\texttt{#1}}\fi
\expandafter\ifx\csname urlprefix\endcsname\relax\def\urlprefix{URL }\fi
\providecommand{\bibinfo}[2]{#2}
\providecommand{\eprint}[2][]{\url{#2}}

\bibitem[{\citenamefont{{El Hanafy} and {Nashed}}(2015)}]{HN15}
\bibinfo{author}{\bibfnamefont{W.}~\bibnamefont{{El Hanafy}}} \bibnamefont{and}
  \bibinfo{author}{\bibfnamefont{G.~G.~L.} \bibnamefont{{Nashed}}},
  \bibinfo{journal}{Eur. Phys. J. C} \textbf{\bibinfo{volume}{75}},
  \bibinfo{pages}{279} (\bibinfo{year}{2015}), \eprint{1409.7199}.

\bibitem[{\citenamefont{{Ade} et~al.}(2014{\natexlab{a}})\citenamefont{{Ade},
  {Aghanim}, {Armitage-Caplan}, {Arnaud}, {Ashdown}, {Atrio-Barandela},
  {Aumont}, {Baccigalupi}, {Banday}, and et~al. {[Planck
  Collaboration]}}}]{Pl1}
\bibinfo{author}{\bibfnamefont{P.~A.~R.} \bibnamefont{{Ade}}},
  \bibinfo{author}{\bibfnamefont{N.}~\bibnamefont{{Aghanim}}},
  \bibinfo{author}{\bibfnamefont{C.}~\bibnamefont{{Armitage-Caplan}}},
  \bibinfo{author}{\bibfnamefont{M.}~\bibnamefont{{Arnaud}}},
  \bibinfo{author}{\bibfnamefont{M.}~\bibnamefont{{Ashdown}}},
  \bibinfo{author}{\bibfnamefont{F.}~\bibnamefont{{Atrio-Barandela}}},
  \bibinfo{author}{\bibfnamefont{J.}~\bibnamefont{{Aumont}}},
  \bibinfo{author}{\bibfnamefont{C.}~\bibnamefont{{Baccigalupi}}},
  \bibinfo{author}{\bibfnamefont{A.~J.} \bibnamefont{{Banday}}},
  \bibnamefont{and} \bibinfo{author}{\bibnamefont{et~al. {[Planck
  Collaboration]}}}, \bibinfo{journal}{Astronomy \& Astrophysics}
  \textbf{\bibinfo{volume}{571}} (\bibinfo{year}{2014}{\natexlab{a}}),
  \bibinfo{note}{arXiv: astro-ph.CO/1303.5076}, \eprint{1303.5076}.

\bibitem[{\citenamefont{{Ade} et~al.}(2014{\natexlab{b}})\citenamefont{{Ade},
  {Aghanim}, {Armitage-Caplan}, {Arnaud}, {Ashdown}, {Atrio-Barandela},
  {Aumont}, {Baccigalupi}, {Banday}, and et~al. {[Planck
  Collaboration]}}}]{Pl2}
\bibinfo{author}{\bibfnamefont{P.~A.~R.} \bibnamefont{{Ade}}},
  \bibinfo{author}{\bibfnamefont{N.}~\bibnamefont{{Aghanim}}},
  \bibinfo{author}{\bibfnamefont{C.}~\bibnamefont{{Armitage-Caplan}}},
  \bibinfo{author}{\bibfnamefont{M.}~\bibnamefont{{Arnaud}}},
  \bibinfo{author}{\bibfnamefont{M.}~\bibnamefont{{Ashdown}}},
  \bibinfo{author}{\bibfnamefont{F.}~\bibnamefont{{Atrio-Barandela}}},
  \bibinfo{author}{\bibfnamefont{J.}~\bibnamefont{{Aumont}}},
  \bibinfo{author}{\bibfnamefont{C.}~\bibnamefont{{Baccigalupi}}},
  \bibinfo{author}{\bibfnamefont{A.~J.} \bibnamefont{{Banday}}},
  \bibnamefont{and} \bibinfo{author}{\bibnamefont{et~al. {[Planck
  Collaboration]}}}, \bibinfo{journal}{Astronomy \& Astrophysics}
  \textbf{\bibinfo{volume}{571}} (\bibinfo{year}{2014}{\natexlab{b}}),
  \bibinfo{note}{arXiv: astro-ph.CO/1303.5082}, \eprint{1303.5082}.

\bibitem[{\citenamefont{{Ade} and et~al. {[BICEP2
  Collaboration]}}(2014)}]{BICEP2}
\bibinfo{author}{\bibfnamefont{P.~A.~R.} \bibnamefont{{Ade}}} \bibnamefont{and}
  \bibinfo{author}{\bibnamefont{et~al. {[BICEP2 Collaboration]}}},
  \bibinfo{journal}{Physical Review Letters} \textbf{\bibinfo{volume}{112}},
  \bibinfo{eid}{241101} (\bibinfo{year}{2014}), \bibinfo{note}{arXiv:
  astro-ph.CO/1403.3985}, \eprint{1403.3985}.

\bibitem[{\citenamefont{{Maluf} et~al.}(2002)\citenamefont{{Maluf}, {da
  Rocha-Neto}, {Tor\'{\i}bio}, and {Castello-Branco}}}]{M2002}
\bibinfo{author}{\bibfnamefont{J.~W.} \bibnamefont{{Maluf}}},
  \bibinfo{author}{\bibfnamefont{J.~F.} \bibnamefont{{da Rocha-Neto}}},
  \bibinfo{author}{\bibfnamefont{T.~M.} \bibnamefont{{Tor\'{\i}bio}}},
  \bibnamefont{and} \bibinfo{author}{\bibfnamefont{K.~H.}
  \bibnamefont{{Castello-Branco}}}, \bibinfo{journal}{Physical Review D}
  \textbf{\bibinfo{volume}{65}}, \bibinfo{eid}{124001} (\bibinfo{year}{2002}),
  \bibinfo{note}{arXiv: gr-qc/0204035}, \eprint{gr-qc/0204035}.

\bibitem[{\citenamefont{{Shirafuji} et~al.}(1996)\citenamefont{{Shirafuji},
  {Nashed}, and {Kobayashi}}}]{nashed:1996}
\bibinfo{author}{\bibfnamefont{T.}~\bibnamefont{{Shirafuji}}},
  \bibinfo{author}{\bibfnamefont{G.~G.~L.} \bibnamefont{{Nashed}}},
  \bibnamefont{and}
  \bibinfo{author}{\bibfnamefont{Y.}~\bibnamefont{{Kobayashi}}},
  \bibinfo{journal}{Progress of Theoretical Physics}
  \textbf{\bibinfo{volume}{96}}, \bibinfo{pages}{933} (\bibinfo{year}{1996}),
  \eprint{gr-qc/9609060}.

\bibitem[{\citenamefont{{Nashed}}(2010)}]{nashed:2010}
\bibinfo{author}{\bibfnamefont{G.~G.~L.} \bibnamefont{{Nashed}}},
  \bibinfo{journal}{Astrophysics and Space Science}
  \textbf{\bibinfo{volume}{330}}, \bibinfo{pages}{173} (\bibinfo{year}{2010}),
  \eprint{1503.01379}.

\bibitem[{\citenamefont{{Nashed}}(2003)}]{nashed:2003}
\bibinfo{author}{\bibfnamefont{G.~G.~L.} \bibnamefont{{Nashed}}},
  \bibinfo{journal}{Chaos Solitons and Fractals} \textbf{\bibinfo{volume}{15}},
  \bibinfo{pages}{841} (\bibinfo{year}{2003}), \eprint{gr-qc/0301008}.

\bibitem[{\citenamefont{{Shirafuji} and {Nashed}}(1997)}]{nashed:1997}
\bibinfo{author}{\bibfnamefont{T.}~\bibnamefont{{Shirafuji}}} \bibnamefont{and}
  \bibinfo{author}{\bibfnamefont{G.~G.~L.} \bibnamefont{{Nashed}}},
  \bibinfo{journal}{Progress of Theoretical Physics}
  \textbf{\bibinfo{volume}{98}}, \bibinfo{pages}{1355} (\bibinfo{year}{1997}),
  \eprint{gr-qc/9711010}.

\bibitem[{\citenamefont{{Nashed}}(2002)}]{nashed:2002}
\bibinfo{author}{\bibfnamefont{G.~G.~L.} \bibnamefont{{Nashed}}},
  \bibinfo{journal}{Nuovo Cimento B Serie} \textbf{\bibinfo{volume}{117}},
  \bibinfo{pages}{521} (\bibinfo{year}{2002}), \eprint{gr-qc/0109017}.

\bibitem[{\citenamefont{{Nashed}}(2006)}]{nashed:2006}
\bibinfo{author}{\bibfnamefont{G.~G.~L.} \bibnamefont{{Nashed}}},
  \bibinfo{journal}{International Journal of Modern Physics A}
  \textbf{\bibinfo{volume}{21}}, \bibinfo{pages}{3181} (\bibinfo{year}{2006}),
  \eprint{gr-qc/0501002}.

\bibitem[{\citenamefont{{Nashed}}(2007)}]{nashed:2007}
\bibinfo{author}{\bibfnamefont{G.~G.~L.} \bibnamefont{{Nashed}}},
  \bibinfo{journal}{European Physical Journal C} \textbf{\bibinfo{volume}{49}},
  \bibinfo{pages}{851} (\bibinfo{year}{2007}), \eprint{0706.0260}.

\bibitem[{\citenamefont{{Wanas}}(2009)}]{Wanas2}
\bibinfo{author}{\bibfnamefont{M.~I.} \bibnamefont{{Wanas}}},
  \bibinfo{journal}{Modern Physics Letters A} \textbf{\bibinfo{volume}{24}},
  \bibinfo{pages}{1749} (\bibinfo{year}{2009}), \eprint{0801.1132}.

\bibitem[{\citenamefont{{Youssef} et~al.}(2008)\citenamefont{{Youssef}, {Abed},
  and {Soleiman}}}]{Nabil1}
\bibinfo{author}{\bibfnamefont{N.~L.} \bibnamefont{{Youssef}}},
  \bibinfo{author}{\bibfnamefont{S.~H.} \bibnamefont{{Abed}}},
  \bibnamefont{and}
  \bibinfo{author}{\bibfnamefont{A.}~\bibnamefont{{Soleiman}}},
  \bibinfo{journal}{ArXiv e-prints}  (\bibinfo{year}{2008}),
  \eprint{0801.3220}.

\bibitem[{\citenamefont{{Youssef} et~al.}(2006)\citenamefont{{Youssef}, {Abed},
  and {Soleiman}}}]{Nabil2}
\bibinfo{author}{\bibfnamefont{N.~L.} \bibnamefont{{Youssef}}},
  \bibinfo{author}{\bibfnamefont{S.~H.} \bibnamefont{{Abed}}},
  \bibnamefont{and}
  \bibinfo{author}{\bibfnamefont{A.}~\bibnamefont{{Soleiman}}},
  \bibinfo{journal}{ArXiv Mathematics e-prints}  (\bibinfo{year}{2006}),
  \eprint{math/0610052}.

\bibitem[{\citenamefont{{Tamim} and {Youssef}}(2006)}]{Nabil3}
\bibinfo{author}{\bibfnamefont{A.~A.} \bibnamefont{{Tamim}}} \bibnamefont{and}
  \bibinfo{author}{\bibfnamefont{N.~L.} \bibnamefont{{Youssef}}},
  \bibinfo{journal}{ArXiv Mathematics e-prints}  (\bibinfo{year}{2006}),
  \eprint{math/0607572}.

\bibitem[{\citenamefont{{Wanas}}(2007)}]{Wanas1}
\bibinfo{author}{\bibfnamefont{M.~I.} \bibnamefont{{Wanas}}},
  \bibinfo{journal}{International Journal of Geometric Methods in Modern
  Physics} \textbf{\bibinfo{volume}{4}}, \bibinfo{pages}{373}
  (\bibinfo{year}{2007}), \eprint{gr-qc/0703036}.

\bibitem[{\citenamefont{{Mikhail} et~al.}(1995)\citenamefont{{Mikhail},
  {Wanas}, and {Eid}}}]{Wanas3}
\bibinfo{author}{\bibfnamefont{F.~I.} \bibnamefont{{Mikhail}}},
  \bibinfo{author}{\bibfnamefont{M.~I.} \bibnamefont{{Wanas}}},
  \bibnamefont{and} \bibinfo{author}{\bibfnamefont{A.~M.} \bibnamefont{{Eid}}},
  \bibinfo{journal}{Astrophysics \& Space Science}
  \textbf{\bibinfo{volume}{228}}, \bibinfo{pages}{221} (\bibinfo{year}{1995}).

\bibitem[{\citenamefont{{Wanas}}(1986)}]{Wanas4}
\bibinfo{author}{\bibfnamefont{M.~I.} \bibnamefont{{Wanas}}},
  \bibinfo{journal}{Astrophysics \& Space Science}
  \textbf{\bibinfo{volume}{127}}, \bibinfo{pages}{21} (\bibinfo{year}{1986}).

\bibitem[{\citenamefont{{Wanas} et~al.}(2014)\citenamefont{{Wanas}, {Youssef},
  and {El Hanafy}}}]{Waleed4}
\bibinfo{author}{\bibfnamefont{M.~I.} \bibnamefont{{Wanas}}},
  \bibinfo{author}{\bibfnamefont{N.~L.} \bibnamefont{{Youssef}}},
  \bibnamefont{and} \bibinfo{author}{\bibfnamefont{W.}~\bibnamefont{{El
  Hanafy}}}, \bibinfo{journal}{ArXiv e-prints}  (\bibinfo{year}{2014}),
  \eprint{1404.2485}.

\bibitem[{\citenamefont{{Youssef} and {Sid-Ahmed}}(2007)}]{NA2007}
\bibinfo{author}{\bibfnamefont{N.~L.} \bibnamefont{{Youssef}}}
  \bibnamefont{and} \bibinfo{author}{\bibfnamefont{A.~M.}
  \bibnamefont{{Sid-Ahmed}}}, \bibinfo{journal}{Reports on Mathematical
  Physics} \textbf{\bibinfo{volume}{60}}, \bibinfo{pages}{39}
  (\bibinfo{year}{2007}), \bibinfo{note}{arXiv: gr-qc/0604111},
  \eprint{0604111}.

\bibitem[{\citenamefont{{Maluf}}(2013)}]{M2013}
\bibinfo{author}{\bibfnamefont{J.~W.} \bibnamefont{{Maluf}}},
  \bibinfo{journal}{Annalen der Physik} \textbf{\bibinfo{volume}{525}},
  \bibinfo{pages}{339} (\bibinfo{year}{2013}), \bibinfo{note}{arXiv:
  gr-qc/1303.3897}, \eprint{1303.3897}.

\bibitem[{\citenamefont{{Li} et~al.}(2011)\citenamefont{{Li}, {Sotiriou}, and
  {Barrow}}}]{1010.1041}
\bibinfo{author}{\bibfnamefont{B.}~\bibnamefont{{Li}}},
  \bibinfo{author}{\bibfnamefont{T.~P.} \bibnamefont{{Sotiriou}}},
  \bibnamefont{and} \bibinfo{author}{\bibfnamefont{J.~D.}
  \bibnamefont{{Barrow}}}, \bibinfo{journal}{Physical Review D}
  \textbf{\bibinfo{volume}{83}}, \bibinfo{eid}{064035} (\bibinfo{year}{2011}),
  \bibinfo{note}{arXiv: gr-qc/1010.1041}, \eprint{1010.1041}.

\bibitem[{\citenamefont{{Sotiriou} et~al.}(2011)\citenamefont{{Sotiriou}, {Li},
  and {Barrow}}}]{1012.4039}
\bibinfo{author}{\bibfnamefont{T.~P.} \bibnamefont{{Sotiriou}}},
  \bibinfo{author}{\bibfnamefont{B.}~\bibnamefont{{Li}}}, \bibnamefont{and}
  \bibinfo{author}{\bibfnamefont{J.~D.} \bibnamefont{{Barrow}}},
  \bibinfo{journal}{Physical Review D} \textbf{\bibinfo{volume}{83}},
  \bibinfo{eid}{104030} (\bibinfo{year}{2011}), \bibinfo{note}{arXiv:
  gr-qc/1012.4039}, \eprint{1012.4039}.

\bibitem[{\citenamefont{{Kr\v{s}\v{s}\'{a}k} and {Saridakis}}(2016)}]{Martin2}
\bibinfo{author}{\bibfnamefont{M.}~\bibnamefont{{Kr\v{s}\v{s}\'{a}k}}}
  \bibnamefont{and} \bibinfo{author}{\bibfnamefont{E.~N.}
  \bibnamefont{{Saridakis}}}, \bibinfo{journal}{Classical and Quantum Gravity}
  \textbf{\bibinfo{volume}{33}}, \bibinfo{eid}{115009} (\bibinfo{year}{2016}),
  \eprint{1510.08432}.

\bibitem[{\citenamefont{{Bengochea} and {Ferraro}}(2009)}]{BF09}
\bibinfo{author}{\bibfnamefont{G.~R.} \bibnamefont{{Bengochea}}}
  \bibnamefont{and}
  \bibinfo{author}{\bibfnamefont{R.}~\bibnamefont{{Ferraro}}},
  \bibinfo{journal}{Physical Review D} \textbf{\bibinfo{volume}{79}},
  \bibinfo{eid}{124019} (\bibinfo{year}{2009}), \bibinfo{note}{arXiv:
  gr-qc/0812.1205}, \eprint{0812.1205}.

\bibitem[{\citenamefont{{Yang}}(2011)}]{Y2011}
\bibinfo{author}{\bibfnamefont{R.-J.} \bibnamefont{{Yang}}},
  \bibinfo{journal}{Europhysics Letters} \textbf{\bibinfo{volume}{93}},
  \bibinfo{pages}{60001} (\bibinfo{year}{2011}), \bibinfo{note}{arXiv:
  gr-qc/1010.1376}, \eprint{1010.1376}.

\bibitem[{\citenamefont{{Cai} et~al.}(2011)\citenamefont{{Cai}, {Chen}, {Dent},
  {Dutta}, and {Saridakis}}}]{CCDDS11}
\bibinfo{author}{\bibfnamefont{Y.-F.} \bibnamefont{{Cai}}},
  \bibinfo{author}{\bibfnamefont{S.-H.} \bibnamefont{{Chen}}},
  \bibinfo{author}{\bibfnamefont{J.~B.} \bibnamefont{{Dent}}},
  \bibinfo{author}{\bibfnamefont{S.}~\bibnamefont{{Dutta}}}, \bibnamefont{and}
  \bibinfo{author}{\bibfnamefont{E.~N.} \bibnamefont{{Saridakis}}},
  \bibinfo{journal}{Classical and Quantum Gravity}
  \textbf{\bibinfo{volume}{28}}, \bibinfo{eid}{215011} (\bibinfo{year}{2011}),
  \bibinfo{note}{arXiv: astro-ph.CO/1104.4349}, \eprint{1104.4349}.

\bibitem[{\citenamefont{{Ferraro} and {Fiorini}}(2011{\natexlab{a}})}]{FF011}
\bibinfo{author}{\bibfnamefont{R.}~\bibnamefont{{Ferraro}}} \bibnamefont{and}
  \bibinfo{author}{\bibfnamefont{F.}~\bibnamefont{{Fiorini}}},
  \bibinfo{journal}{Physical Review D} \textbf{\bibinfo{volume}{84}},
  \bibinfo{eid}{083518} (\bibinfo{year}{2011}{\natexlab{a}}),
  \bibinfo{note}{arXiv: gr-qc/1109.4209}, \eprint{1109.4209}.

\bibitem[{\citenamefont{{Ferraro} and {Fiorini}}(2011{\natexlab{b}})}]{FF11}
\bibinfo{author}{\bibfnamefont{R.}~\bibnamefont{{Ferraro}}} \bibnamefont{and}
  \bibinfo{author}{\bibfnamefont{F.}~\bibnamefont{{Fiorini}}},
  \bibinfo{journal}{Physics Letters B} \textbf{\bibinfo{volume}{702}},
  \bibinfo{pages}{75} (\bibinfo{year}{2011}{\natexlab{b}}),
  \bibinfo{note}{arXiv: gr-qc/1103.0824}, \eprint{1103.0824}.

\bibitem[{\citenamefont{{Iorio} and {Saridakis}}(2012)}]{IS12}
\bibinfo{author}{\bibfnamefont{L.}~\bibnamefont{{Iorio}}} \bibnamefont{and}
  \bibinfo{author}{\bibfnamefont{E.~N.} \bibnamefont{{Saridakis}}},
  \bibinfo{journal}{Monthly Notices of Royal Astronomical Society}
  \textbf{\bibinfo{volume}{427}}, \bibinfo{pages}{1555} (\bibinfo{year}{2012}),
  \bibinfo{note}{arXiv: gr-qc/1203.5781}, \eprint{1203.5781}.

\bibitem[{\citenamefont{{Capozziello} et~al.}(2013)\citenamefont{{Capozziello},
  {Gonz{\'a}lez}, {Saridakis}, and {V{\'a}squez}}}]{CGS13}
\bibinfo{author}{\bibfnamefont{S.}~\bibnamefont{{Capozziello}}},
  \bibinfo{author}{\bibfnamefont{P.~A.} \bibnamefont{{Gonz{\'a}lez}}},
  \bibinfo{author}{\bibfnamefont{E.~N.} \bibnamefont{{Saridakis}}},
  \bibnamefont{and}
  \bibinfo{author}{\bibfnamefont{Y.}~\bibnamefont{{V{\'a}squez}}},
  \bibinfo{journal}{Journal of High Energy Physics}
  \textbf{\bibinfo{volume}{2}}, \bibinfo{eid}{39} (\bibinfo{year}{2013}),
  \bibinfo{note}{arXiv: hep-th/1201.1098}, \eprint{1210.1098}.

\bibitem[{\citenamefont{{Nashed}}(2013{\natexlab{a}})}]{Nashed1}
\bibinfo{author}{\bibfnamefont{G.~G.~L.} \bibnamefont{{Nashed}}},
  \bibinfo{journal}{Physical Review D} \textbf{\bibinfo{volume}{88}},
  \bibinfo{eid}{104034} (\bibinfo{year}{2013}{\natexlab{a}}),
  \bibinfo{note}{arXiv: gr-qc/1311.3131}, \eprint{1311.3131}.

\bibitem[{\citenamefont{{Nashed}}(2013{\natexlab{b}})}]{Nashed2}
\bibinfo{author}{\bibfnamefont{G.~G.~L.} \bibnamefont{{Nashed}}},
  \bibinfo{journal}{General Relativity and Gravitation}
  \textbf{\bibinfo{volume}{45}}, \bibinfo{pages}{1887}
  (\bibinfo{year}{2013}{\natexlab{b}}), \bibinfo{note}{arXiv:
  gr-qc/1502.05219}, \eprint{1502.05219}.

\bibitem[{\citenamefont{{Rodrigues} et~al.}(2013)\citenamefont{{Rodrigues},
  {Houndjo}, {Tossa}, {Momeni}, and {Myrzakulov}}}]{RHTMM13}
\bibinfo{author}{\bibfnamefont{M.~E.} \bibnamefont{{Rodrigues}}},
  \bibinfo{author}{\bibfnamefont{M.~J.~S.} \bibnamefont{{Houndjo}}},
  \bibinfo{author}{\bibfnamefont{J.}~\bibnamefont{{Tossa}}},
  \bibinfo{author}{\bibfnamefont{D.}~\bibnamefont{{Momeni}}}, \bibnamefont{and}
  \bibinfo{author}{\bibfnamefont{R.}~\bibnamefont{{Myrzakulov}}},
  \bibinfo{journal}{Journal of Cosmology and Astroparticle Physics}
  \textbf{\bibinfo{volume}{11}}, \bibinfo{eid}{024} (\bibinfo{year}{2013}),
  \bibinfo{note}{arXiv: gr-qc/1306.2280}, \eprint{1306.2280}.

\bibitem[{\citenamefont{{Nashed}}(2014)}]{Nashed3}
\bibinfo{author}{\bibfnamefont{G.~G.~L.} \bibnamefont{{Nashed}}},
  \bibinfo{journal}{Europhysics Letters} \textbf{\bibinfo{volume}{105}},
  \bibinfo{eid}{10001} (\bibinfo{year}{2014}), \bibinfo{note}{arXiv:
  gr-qc/1501.00974}, \eprint{1501.00974}.

\bibitem[{\citenamefont{{Bejarano} et~al.}(2015)\citenamefont{{Bejarano},
  {Ferraro}, and {Guzm{\'a}n}}}]{BFG15}
\bibinfo{author}{\bibfnamefont{C.}~\bibnamefont{{Bejarano}}},
  \bibinfo{author}{\bibfnamefont{R.}~\bibnamefont{{Ferraro}}},
  \bibnamefont{and} \bibinfo{author}{\bibfnamefont{M.~J.}
  \bibnamefont{{Guzm{\'a}n}}}, \bibinfo{journal}{The European Physical Journal
  C} \textbf{\bibinfo{volume}{75}}, \bibinfo{pages}{77} (\bibinfo{year}{2015}),
  \bibinfo{note}{arXiv: gr-qc/1412.0641}, \eprint{1412.0641}.

\bibitem[{\citenamefont{{Nashed}}(2015{\natexlab{a}})}]{Nashed4}
\bibinfo{author}{\bibfnamefont{G.~G.~L.} \bibnamefont{{Nashed}}},
  \bibinfo{journal}{Journal of the Physical Society of Japan}
  \textbf{\bibinfo{volume}{84}}, \bibinfo{eid}{044006}
  (\bibinfo{year}{2015}{\natexlab{a}}).

\bibitem[{\citenamefont{{Nashed}}(2015{\natexlab{b}})}]{Nashed5}
\bibinfo{author}{\bibfnamefont{G.~G.~L.} \bibnamefont{{Nashed}}},
  \bibinfo{journal}{International Journal of Modern Physics D}
  \textbf{\bibinfo{volume}{24}}, \bibinfo{eid}{1550007}
  (\bibinfo{year}{2015}{\natexlab{b}}).

\bibitem[{\citenamefont{{El Hanafy} and
  {Nashed}}(2016{\natexlab{a}})}]{Hanafy:2016}
\bibinfo{author}{\bibfnamefont{W.}~\bibnamefont{{El Hanafy}}} \bibnamefont{and}
  \bibinfo{author}{\bibfnamefont{G.~G.~L.} \bibnamefont{{Nashed}}},
  \bibinfo{journal}{Astrophys. Space Sci.} \textbf{\bibinfo{volume}{361}},
  \bibinfo{eid}{68} (\bibinfo{year}{2016}{\natexlab{a}}), \eprint{1507.07377}.

\bibitem[{\citenamefont{{Bamba} et~al.}(2012)\citenamefont{{Bamba},
  {Capozziello}, {Nojiri}, and {Odintsov}}}]{1205.3421}
\bibinfo{author}{\bibfnamefont{K.}~\bibnamefont{{Bamba}}},
  \bibinfo{author}{\bibfnamefont{S.}~\bibnamefont{{Capozziello}}},
  \bibinfo{author}{\bibfnamefont{S.}~\bibnamefont{{Nojiri}}}, \bibnamefont{and}
  \bibinfo{author}{\bibfnamefont{S.~D.} \bibnamefont{{Odintsov}}},
  \bibinfo{journal}{Astrophysics and Space Science}
  \textbf{\bibinfo{volume}{342}}, \bibinfo{pages}{155} (\bibinfo{year}{2012}),
  \bibinfo{note}{arXiv: gr-qc/1205.3421}, \eprint{1205.3421}.

\bibitem[{\citenamefont{{Nashed}}(2011)}]{Nashed:2011}
\bibinfo{author}{\bibfnamefont{G.~G.~L.} \bibnamefont{{Nashed}}},
  \bibinfo{journal}{ArXiv e-prints}  (\bibinfo{year}{2011}),
  \eprint{1111.0003}.

\bibitem[{\citenamefont{{Momeni} and
  {Myrzakulov}}(2014{\natexlab{a}})}]{Momeni:2014a}
\bibinfo{author}{\bibfnamefont{D.}~\bibnamefont{{Momeni}}} \bibnamefont{and}
  \bibinfo{author}{\bibfnamefont{R.}~\bibnamefont{{Myrzakulov}}},
  \bibinfo{journal}{Int. J. Geom. Meth. Mod. Phys.}
  \textbf{\bibinfo{volume}{11}}, \bibinfo{pages}{1450077}
  (\bibinfo{year}{2014}{\natexlab{a}}), \eprint{1405.5863}.

\bibitem[{\citenamefont{{Momeni} and
  {Myrzakulov}}(2014{\natexlab{b}})}]{Momeni:2014b}
\bibinfo{author}{\bibfnamefont{D.}~\bibnamefont{{Momeni}}} \bibnamefont{and}
  \bibinfo{author}{\bibfnamefont{R.}~\bibnamefont{{Myrzakulov}}},
  \bibinfo{journal}{European Physical Journal Plus}
  \textbf{\bibinfo{volume}{129}}, \bibinfo{eid}{137}
  (\bibinfo{year}{2014}{\natexlab{b}}), \eprint{1404.0778}.

\bibitem[{\citenamefont{{Bamba} et~al.}(2014)\citenamefont{{Bamba}, {Nojiri},
  and {Odintsov}}}]{BNO14}
\bibinfo{author}{\bibfnamefont{K.}~\bibnamefont{{Bamba}}},
  \bibinfo{author}{\bibfnamefont{S.}~\bibnamefont{{Nojiri}}}, \bibnamefont{and}
  \bibinfo{author}{\bibfnamefont{S.~D.} \bibnamefont{{Odintsov}}},
  \bibinfo{journal}{Physics Letters B} \textbf{\bibinfo{volume}{731}},
  \bibinfo{pages}{257} (\bibinfo{year}{2014}), \bibinfo{note}{arXiv:
  gr-qc/1401.7378}, \eprint{1401.7378}.

\bibitem[{\citenamefont{{Bamba} and {Odintsov}}(2014)}]{BO14}
\bibinfo{author}{\bibfnamefont{K.}~\bibnamefont{{Bamba}}} \bibnamefont{and}
  \bibinfo{author}{\bibfnamefont{S.~D.} \bibnamefont{{Odintsov}}}
  (\bibinfo{year}{2014}), \bibinfo{note}{arXiv: hep-th/1402.7114},
  \eprint{1402.7114}.

\bibitem[{\citenamefont{{Jamil} et~al.}(2014)\citenamefont{{Jamil}, {Momeni},
  and {Myrzakulov}}}]{JMM14}
\bibinfo{author}{\bibfnamefont{M.}~\bibnamefont{{Jamil}}},
  \bibinfo{author}{\bibfnamefont{D.}~\bibnamefont{{Momeni}}}, \bibnamefont{and}
  \bibinfo{author}{\bibfnamefont{R.}~\bibnamefont{{Myrzakulov}}},
  \bibinfo{journal}{International Journal of Theoretical Physics}
  (\bibinfo{year}{2014}), \bibinfo{note}{arXiv: gr-qc/1309.3269},
  \eprint{1309.3269}.

\bibitem[{\citenamefont{{Harko} et~al.}(2014)\citenamefont{{Harko}, {Lobo},
  {Otalora}, and {Saridakis}}}]{HLOS14}
\bibinfo{author}{\bibfnamefont{T.}~\bibnamefont{{Harko}}},
  \bibinfo{author}{\bibfnamefont{F.~S.~N.} \bibnamefont{{Lobo}}},
  \bibinfo{author}{\bibfnamefont{G.}~\bibnamefont{{Otalora}}},
  \bibnamefont{and} \bibinfo{author}{\bibfnamefont{E.~N.}
  \bibnamefont{{Saridakis}}}, \bibinfo{journal}{Physical Review D}
  \textbf{\bibinfo{volume}{89}}, \bibinfo{eid}{124036} (\bibinfo{year}{2014}),
  \bibinfo{note}{arXiv: gr-qc/1404.6212}, \eprint{1404.6212}.

\bibitem[{\citenamefont{{Nashed} and {El Hanafy}}(2014)}]{NH14}
\bibinfo{author}{\bibfnamefont{G.~G.~L.} \bibnamefont{{Nashed}}}
  \bibnamefont{and} \bibinfo{author}{\bibfnamefont{W.}~\bibnamefont{{El
  Hanafy}}}, \bibinfo{journal}{The European Physical Journal C}
  \textbf{\bibinfo{volume}{74}}, \bibinfo{pages}{3099} (\bibinfo{year}{2014}),
  \bibinfo{note}{arXiv: gr-qc/1403.0913}, \eprint{1403.0913}.

\bibitem[{\citenamefont{{Wanas} and {Hassan}}(2014)}]{WH14}
\bibinfo{author}{\bibfnamefont{M.~I.} \bibnamefont{{Wanas}}} \bibnamefont{and}
  \bibinfo{author}{\bibfnamefont{H.~A.} \bibnamefont{{Hassan}}},
  \bibinfo{journal}{International Journal of Theoretical Physics}
  \textbf{\bibinfo{volume}{53}}, \bibinfo{pages}{3901} (\bibinfo{year}{2014}).

\bibitem[{\citenamefont{{Wu} et~al.}(2015)\citenamefont{{Wu}, {Chen}, {Wang},
  and {Wei}}}]{1503.05281}
\bibinfo{author}{\bibfnamefont{Y.}~\bibnamefont{{Wu}}},
  \bibinfo{author}{\bibfnamefont{Z.-C.} \bibnamefont{{Chen}}},
  \bibinfo{author}{\bibfnamefont{J.}~\bibnamefont{{Wang}}}, \bibnamefont{and}
  \bibinfo{author}{\bibfnamefont{H.}~\bibnamefont{{Wei}}},
  \bibinfo{journal}{Communications in Theoretical Physics}
  \textbf{\bibinfo{volume}{63}}, \bibinfo{pages}{701} (\bibinfo{year}{2015}),
  \bibinfo{note}{arXiv: gr-qc/1503.05281}, \eprint{1503.05281}.

\bibitem[{\citenamefont{{Junior} et~al.}(2015)\citenamefont{{Junior},
  {Rodrigues}, and {Houndjo}}}]{1503.07427}
\bibinfo{author}{\bibfnamefont{E.~L.~B.} \bibnamefont{{Junior}}},
  \bibinfo{author}{\bibfnamefont{M.~E.} \bibnamefont{{Rodrigues}}},
  \bibnamefont{and} \bibinfo{author}{\bibfnamefont{M.~J.~S.}
  \bibnamefont{{Houndjo}}}, \bibinfo{journal}{Journal of Cosmology and
  Astroparticle Physics} \textbf{\bibinfo{volume}{6}}, \bibinfo{eid}{037}
  (\bibinfo{year}{2015}), \bibinfo{note}{arXiv: gr-qc/1503.07427},
  \eprint{1503.07427}.

\bibitem[{\citenamefont{{El Hanafy} and
  {Nashed}}(2016{\natexlab{b}})}]{Waleed:2016}
\bibinfo{author}{\bibfnamefont{W.}~\bibnamefont{{El Hanafy}}} \bibnamefont{and}
  \bibinfo{author}{\bibfnamefont{G.~G.~L.} \bibnamefont{{Nashed}}},
  \bibinfo{journal}{Astrophys. Space Sci.} \textbf{\bibinfo{volume}{361}},
  \bibinfo{pages}{197} (\bibinfo{year}{2016}{\natexlab{b}}),
  \eprint{1410.2467}.

\bibitem[{\citenamefont{{Nunes} et~al.}(2016)\citenamefont{{Nunes}, {Pan}, and
  {Saridakis}}}]{Nunes:2016}
\bibinfo{author}{\bibfnamefont{R.~C.} \bibnamefont{{Nunes}}},
  \bibinfo{author}{\bibfnamefont{S.}~\bibnamefont{{Pan}}}, \bibnamefont{and}
  \bibinfo{author}{\bibfnamefont{E.~N.} \bibnamefont{{Saridakis}}},
  \bibinfo{journal}{ArXiv e-prints}  (\bibinfo{year}{2016}),
  \eprint{1606.04359}.

\bibitem[{\citenamefont{{Bamba} et~al.}(2016)\citenamefont{{Bamba}, {Odintsov},
  and {Saridakis}}}]{BambaOS:2016}
\bibinfo{author}{\bibfnamefont{K.}~\bibnamefont{{Bamba}}},
  \bibinfo{author}{\bibfnamefont{S.~D.} \bibnamefont{{Odintsov}}},
  \bibnamefont{and} \bibinfo{author}{\bibfnamefont{E.~N.}
  \bibnamefont{{Saridakis}}}, \bibinfo{journal}{ArXiv e-prints}
  (\bibinfo{year}{2016}), \eprint{1605.02461}.

\bibitem[{\citenamefont{{Otalora} and {Saridakis}}(2016)}]{OtaloraS:2016}
\bibinfo{author}{\bibfnamefont{G.}~\bibnamefont{{Otalora}}} \bibnamefont{and}
  \bibinfo{author}{\bibfnamefont{E.~N.} \bibnamefont{{Saridakis}}},
  \bibinfo{journal}{ArXiv e-prints}  (\bibinfo{year}{2016}),
  \eprint{1605.04599}.

\bibitem[{\citenamefont{{Cai} et~al.}(2014)\citenamefont{{Cai}, {Quintin},
  {Saridakis}, and {Wilson-Ewing}}}]{CQSW14}
\bibinfo{author}{\bibfnamefont{Y.-F.} \bibnamefont{{Cai}}},
  \bibinfo{author}{\bibfnamefont{J.}~\bibnamefont{{Quintin}}},
  \bibinfo{author}{\bibfnamefont{E.~N.} \bibnamefont{{Saridakis}}},
  \bibnamefont{and}
  \bibinfo{author}{\bibfnamefont{E.}~\bibnamefont{{Wilson-Ewing}}},
  \bibinfo{journal}{Journal of Cosmology and Astroparticle Physics}
  \textbf{\bibinfo{volume}{7}}, \bibinfo{eid}{033} (\bibinfo{year}{2014}),
  \bibinfo{note}{arXiv: astro-ph/1404.4364}, \eprint{1404.4364}.

\bibitem[{\citenamefont{{Haro} and {Amor{\'o}s}}(2014)}]{Haro:2014}
\bibinfo{author}{\bibfnamefont{J.}~\bibnamefont{{Haro}}} \bibnamefont{and}
  \bibinfo{author}{\bibfnamefont{J.}~\bibnamefont{{Amor{\'o}s}}},
  \bibinfo{journal}{Journal of Cosmology and Astroparticle Physics}
  \textbf{\bibinfo{volume}{12}}, \bibinfo{eid}{031} (\bibinfo{year}{2014}),
  \eprint{1406.0369}.

\bibitem[{\citenamefont{Bamba et~al.}(2016)\citenamefont{Bamba, Nashed, Hanafy,
  and Ibrahim}}]{Bamba:2016}
\bibinfo{author}{\bibfnamefont{K.}~\bibnamefont{Bamba}},
  \bibinfo{author}{\bibfnamefont{G.~G.~L.} \bibnamefont{Nashed}},
  \bibinfo{author}{\bibfnamefont{W.~E.} \bibnamefont{Hanafy}},
  \bibnamefont{and} \bibinfo{author}{\bibfnamefont{S.~K.}
  \bibnamefont{Ibrahim}} (\bibinfo{year}{2016}), \eprint{1604.07604}.

\bibitem[{\citenamefont{Cai et~al.}(2015)\citenamefont{Cai, Capozziello,
  De~Laurentis, and Saridakis}}]{Saridakis1}
\bibinfo{author}{\bibfnamefont{Y.-F.} \bibnamefont{Cai}},
  \bibinfo{author}{\bibfnamefont{S.}~\bibnamefont{Capozziello}},
  \bibinfo{author}{\bibfnamefont{M.}~\bibnamefont{De~Laurentis}},
  \bibnamefont{and} \bibinfo{author}{\bibfnamefont{E.~N.}
  \bibnamefont{Saridakis}} (\bibinfo{year}{2015}), \eprint{1511.07586}.

\bibitem[{\citenamefont{{Robertson}}(1932)}]{R32}
\bibinfo{author}{\bibfnamefont{H.~P.} \bibnamefont{{Robertson}}},
  \bibinfo{journal}{Annals of Mathematics} \textbf{\bibinfo{volume}{33}},
  \bibinfo{pages}{496} (\bibinfo{year}{1932}).

\bibitem[{\citenamefont{{Linder}}(2010)}]{L10}
\bibinfo{author}{\bibfnamefont{E.~V.} \bibnamefont{{Linder}}},
  \bibinfo{journal}{Physical Review D} \textbf{\bibinfo{volume}{81}},
  \bibinfo{eid}{127301} (\bibinfo{year}{2010}), \bibinfo{note}{arXiv:
  astro-ph.CO/1005.3039}, \eprint{1005.3039}.

\bibitem[{\citenamefont{{Bamba} et~al.}(2011)\citenamefont{{Bamba}, {Geng},
  {Lee}, and {Luo}}}]{1011.0508}
\bibinfo{author}{\bibfnamefont{K.}~\bibnamefont{{Bamba}}},
  \bibinfo{author}{\bibfnamefont{C.-Q.} \bibnamefont{{Geng}}},
  \bibinfo{author}{\bibfnamefont{C.-C.} \bibnamefont{{Lee}}}, \bibnamefont{and}
  \bibinfo{author}{\bibfnamefont{L.-W.} \bibnamefont{{Luo}}},
  \bibinfo{journal}{Journal of Cosmology and Astroparticle Physics}
  \textbf{\bibinfo{volume}{1}}, \bibinfo{eid}{021} (\bibinfo{year}{2011}),
  \bibinfo{note}{arXiv: astro-ph.CO/1011.0508}, \eprint{1011.0508}.

\bibitem[{\citenamefont{{Setare} et~al.}(2016)\citenamefont{{Setare}, {Momeni},
  {Kamali}, and {Myrzakulov}}}]{Setare:2016}
\bibinfo{author}{\bibfnamefont{M.~R.} \bibnamefont{{Setare}}},
  \bibinfo{author}{\bibfnamefont{D.}~\bibnamefont{{Momeni}}},
  \bibinfo{author}{\bibfnamefont{V.}~\bibnamefont{{Kamali}}}, \bibnamefont{and}
  \bibinfo{author}{\bibfnamefont{R.}~\bibnamefont{{Myrzakulov}}},
  \bibinfo{journal}{International Journal of Theoretical Physics}
  \textbf{\bibinfo{volume}{55}}, \bibinfo{pages}{1003} (\bibinfo{year}{2016}),
  \eprint{1409.3200}.

\bibitem[{\citenamefont{{Liddle} and {Lyth}}(2000)}]{LL2000}
\bibinfo{author}{\bibfnamefont{A.~R.} \bibnamefont{{Liddle}}} \bibnamefont{and}
  \bibinfo{author}{\bibfnamefont{D.~H.} \bibnamefont{{Lyth}}},
  \emph{\bibinfo{title}{Cosmological inflation and large-scale structure}}
  (\bibinfo{publisher}{Cambridge University Press}, \bibinfo{year}{2000}).

\bibitem[{\citenamefont{{Bousso} et~al.}(2015)\citenamefont{{Bousso}, {Harlow},
  and {Senatore}}}]{BHS13}
\bibinfo{author}{\bibfnamefont{R.}~\bibnamefont{{Bousso}}},
  \bibinfo{author}{\bibfnamefont{D.}~\bibnamefont{{Harlow}}}, \bibnamefont{and}
  \bibinfo{author}{\bibfnamefont{L.}~\bibnamefont{{Senatore}}},
  \bibinfo{journal}{Physical Review D} \textbf{\bibinfo{volume}{91}},
  \bibinfo{pages}{083527} (\bibinfo{year}{2015}), \bibinfo{note}{arXiv:
  hep-th/13.09.4060}, \eprint{1309.4060}.

\bibitem[{\citenamefont{{Kallosh} et~al.}(2014)\citenamefont{{Kallosh},
  {Linde}, and {Roest}}}]{KLR214}
\bibinfo{author}{\bibfnamefont{R.}~\bibnamefont{{Kallosh}}},
  \bibinfo{author}{\bibfnamefont{A.}~\bibnamefont{{Linde}}}, \bibnamefont{and}
  \bibinfo{author}{\bibfnamefont{D.}~\bibnamefont{{Roest}}},
  \bibinfo{journal}{JHEP} \textbf{\bibinfo{volume}{8}}, \bibinfo{eid}{52}
  (\bibinfo{year}{2014}), \eprint{1405.3646}.

\bibitem[{\citenamefont{{Lucchin} and {Matarrese}}(1985)}]{LM1985}
\bibinfo{author}{\bibfnamefont{F.}~\bibnamefont{{Lucchin}}} \bibnamefont{and}
  \bibinfo{author}{\bibfnamefont{S.}~\bibnamefont{{Matarrese}}},
  \bibinfo{journal}{Phys. Rev. D} \textbf{\bibinfo{volume}{32}},
  \bibinfo{pages}{1316} (\bibinfo{year}{1985}).

\bibitem[{\citenamefont{Corless et~al.}(1996)\citenamefont{Corless, Gonnet,
  Hare, Jeffrey, and Knuth}}]{corless1996lambertw}
\bibinfo{author}{\bibfnamefont{R.~M.} \bibnamefont{Corless}},
  \bibinfo{author}{\bibfnamefont{G.~H.} \bibnamefont{Gonnet}},
  \bibinfo{author}{\bibfnamefont{D.~E.} \bibnamefont{Hare}},
  \bibinfo{author}{\bibfnamefont{D.~J.} \bibnamefont{Jeffrey}},
  \bibnamefont{and} \bibinfo{author}{\bibfnamefont{D.~E.} \bibnamefont{Knuth}},
  \bibinfo{journal}{Advances in Computational mathematics}
  \textbf{\bibinfo{volume}{5}}, \bibinfo{pages}{329} (\bibinfo{year}{1996}).

\end{thebibliography}

%%%%%%%%%%%%%%%%%%%%%%%%%%%%%%%%%%%%%%%%%%%%%%%%%%%%%%%%%%%%%%%%%%%%%%%%%%%%%%%%%%%%%%
\end{document}